\documentstyle[twoside]{article}

\catcode`\@=11
\long\def\@makefntext#1{
\protect\noindent \hbox to 3.2pt {\hskip-.9pt  
$^{{\eightrm\@thefnmark}}$\hfil}#1\hfill}               

\def\thefootnote{\fnsymbol{footnote}}
\def\@makefnmark{\hbox to 0pt{$^{\@thefnmark}$\hss}}    
        
\def\ps@myheadings{\let\@mkboth\@gobbletwo
\def\@oddhead{\hbox{}
\rightmark\hfil\eightrm\thepage}   
\def\@oddfoot{}\def\@evenhead{\eightrm\thepage\hfil
\leftmark\hbox{}}\def\@evenfoot{}
\def\sectionmark##1{}\def\subsectionmark##1{}}

\oddsidemargin=\evensidemargin
\renewcommand{\thefootnote}{\fnsymbol{footnote}}

\newcounter{sectionc}\newcounter{subsectionc}\newcounter{subsubsectionc}
\renewcommand{\section}[1] {\vspace{12pt}\addtocounter{sectionc}{1} 
\setcounter{subsectionc}{0}\setcounter{subsubsectionc}{0}\noindent 
        {\tenbf\thesectionc. #1}\par\vspace{5pt}}
\renewcommand{\subsection}[1] {\vspace{12pt}\addtocounter{subsectionc}{1} 
        \setcounter{subsubsectionc}{0}\noindent 
        {\bf\thesectionc.\thesubsectionc. {\kern1pt \bfit #1}}\par\vspace{5pt}}
\renewcommand{\subsubsection}[1] {\vspace{12pt}\addtocounter{subsubsectionc}{1}
        \noindent{\tenrm\thesectionc.\thesubsectionc.\thesubsubsectionc.
        {\kern1pt \tenit #1}}\par\vspace{5pt}}
\newcommand{\nonumsection}[1] {\vspace{12pt}\noindent{\tenbf #1}
        \par\vspace{5pt}}

\newcounter{appendixc}
\newcounter{subappendixc}[appendixc]
\newcounter{subsubappendixc}[subappendixc]
\renewcommand{\thesubappendixc}{\Alph{appendixc}.\arabic{subappendixc}}
\renewcommand{\thesubsubappendixc}
        {\Alph{appendixc}.\arabic{subappendixc}.\arabic{subsubappendixc}}

\renewcommand{\appendix}[1] {\vspace{12pt}
        \refstepcounter{appendixc}
        \setcounter{figure}{0}
        \setcounter{table}{0}
        \setcounter{lemma}{0}
        \setcounter{theorem}{0}
        \setcounter{corollary}{0}
        \setcounter{definition}{0}
        \setcounter{equation}{0}
        \renewcommand{\thefigure}{\Alph{appendixc}.\arabic{figure}}
        \renewcommand{\thetable}{\Alph{appendixc}.\arabic{table}}
        \renewcommand{\theappendixc}{\Alph{appendixc}}
        \renewcommand{\thelemma}{\Alph{appendixc}.\arabic{lemma}}
        \renewcommand{\thetheorem}{\Alph{appendixc}.\arabic{theorem}}
        \renewcommand{\thedefinition}{\Alph{appendixc}.\arabic{definition}}
        \renewcommand{\thecorollary}{\Alph{appendixc}.\arabic{corollary}}
        \renewcommand{\theequation}{\Alph{appendixc}.\arabic{equation}}
        \noindent{\tenbf Appendix \theappendixc #1}\par\vspace{5pt}}
\newcommand{\subappendix}[1] {\vspace{12pt}
        \refstepcounter{subappendixc}
        \noindent{\bf Appendix \thesubappendixc. {\kern1pt \bfit #1}}
        \par\vspace{5pt}}
\newcommand{\subsubappendix}[1] {\vspace{12pt}
        \refstepcounter{subsubappendixc}
        \noindent{\rm Appendix \thesubsubappendixc. {\kern1pt \tenit #1}}
        \par\vspace{5pt}}

\newcommand{\textlineskip}{\baselineskip=13pt}
\newcommand{\smalllineskip}{\baselineskip=10pt}

\def\eightcirc{
\begin{picture}(0,0)
\put(4.4,1.8){\circle{6.5}}
\end{picture}}
\def\eightcopyright{\eightcirc\kern2.7pt\hbox{\eightrm c}} 

\newcommand{\copyrightheading}[1]
        {\vspace*{-2.5cm}\smalllineskip{\flushleft
        {\footnotesize International Journal of Modern Physics A, #1}\\
        {\footnotesize $\eightcopyright$\, World Scientific Publishing
         Company}\\
         }}

\newcommand{\publisher}[2]{{\begin{center}\footnotesize\smalllineskip 
        Received #1\\
        Revised #2
        \end{center}
        }}

\def\abstracts#1#2#3{{
        \centering{\begin{minipage}{4.5in}\baselineskip=10pt\footnotesize
        \parindent=0pt #1\par 
        \parindent=15pt #2\par
        \parindent=15pt #3
        \end{minipage}}\par}} 

\renewenvironment{thebibliography}[1]
        {\frenchspacing
         \ninerm\baselineskip=11pt
         \begin{list}{\arabic{enumi}.}
        {\usecounter{enumi}\setlength{\parsep}{0pt}
         \setlength{\leftmargin 12.7pt}{\rightmargin 0pt} 
         \setlength{\itemsep}{0pt} \settowidth
        {\labelwidth}{#1.}\sloppy}}{\end{list}}

\newcounter{itemlistc}
\newcounter{romanlistc}
\newcounter{alphlistc}
\newcounter{arabiclistc}

\newcommand{\fcaption}[1]{
        \refstepcounter{figure}
        \setbox\@tempboxa = \hbox{\footnotesize Fig.~\thefigure. #1}
        \ifdim \wd\@tempboxa > 5in
           {\begin{center}
        \parbox{5in}{\footnotesize\smalllineskip Fig.~\thefigure. #1}
            \end{center}}
        \else
             {\begin{center}
             {\footnotesize Fig.~\thefigure. #1}
              \end{center}}
        \fi}

\newcommand{\tcaption}[1]{
        \refstepcounter{table}
        \setbox\@tempboxa = \hbox{\footnotesize Table~\thetable. #1}
        \ifdim \wd\@tempboxa > 5in
           {\begin{center}
        \parbox{5in}{\footnotesize\smalllineskip Table~\thetable. #1}
            \end{center}}
        \else
             {\begin{center}
             {\footnotesize Table~\thetable. #1}
              \end{center}}
        \fi}

\def\@citex[#1]#2{\if@filesw\immediate\write\@auxout
        {\string\citation{#2}}\fi
\def\@citea{}\@cite{\@for\@citeb:=#2\do
        {\@citea\def\@citea{,}\@ifundefined
        {b@\@citeb}{{\bf ?}\@warning
        {Citation `\@citeb' on page \thepage \space undefined}}
        {\csname b@\@citeb\endcsname}}}{#1}}

\newif\if@cghi
\def\cite{\@cghitrue\@ifnextchar [{\@tempswatrue
        \@citex}{\@tempswafalse\@citex[]}}
\def\citelow{\@cghifalse\@ifnextchar [{\@tempswatrue
        \@citex}{\@tempswafalse\@citex[]}}
\def\@cite#1#2{{$\null^{#1}$\if@tempswa\typeout
        {IJCGA warning: optional citation argument 
        ignored: `#2'} \fi}}

\def\pmb#1{\setbox0=\hbox{#1}
        \kern-.025em\copy0\kern-\wd0
        \kern.05em\copy0\kern-\wd0
        \kern-.025em\raise.0433em\box0}

\def\fnt#1#2{\footnotetext{\kern-.3em
        {$^{\mbox{\scriptsize #1}}$}{#2}}}

\def\fpage#1{\begingroup
\voffset=.3in
\thispagestyle{empty}\begin{table}[b]\centerline{\footnotesize #1}
        \end{table}\endgroup}

\def\runninghead#1#2{\pagestyle{myheadings}
\markboth{{\protect\footnotesize\it{\quad #1}}\hfill}
{\hfill{\protect\footnotesize\it{#2\quad}}}}
\font\tenrm=cmr10
\font\tenit=cmti10 
\font\tenbf=cmbx10
\font\bfit=cmbxti10 at 10pt
\font\ninerm=cmr9

\font\eightrm=cmr8

\def\qed{\hbox{${\vcenter{\vbox{                        
   \hrule height 0.4pt\hbox{\vrule width 0.4pt height 6pt
   \kern5pt\vrule width 0.4pt}\hrule height 0.4pt}}}$}}

\renewcommand{\thefootnote}{\fnsymbol{footnote}}        

\begin{document}

\runninghead{SUBDYNAMICS OF RELEVANT OBSERVABLES} {SUBDYNAMICS OF
  RELEVANT OBSERVABLES} 

\normalsize\textlineskip
\thispagestyle{empty}
\setcounter{page}{1}

\copyrightheading{}                     

\vspace*{0.88truein}

\fpage{1}
\centerline{\bf SUBDYNAMICS OF RELEVANT OBSERVABLES:}
\vspace*{0.035truein}
\centerline{\bf A FIELD THEORETICAL APPROACH}
\vspace*{0.37truein}
\centerline{\footnotesize Ludovico Lanz}
\vspace*{0.015truein}
\centerline{\footnotesize\it Dipartimento di Fisica dell'Universit\`a
  di Milano and INFN,} 
\baselineskip=10pt
\centerline{\footnotesize\it Sezione di Milano,}
\baselineskip=10pt
\centerline{\footnotesize\it 
Via Celoria 16, I-20133 Milan,
Italy}
\baselineskip=10pt
\centerline{\footnotesize\it E-mail: lanz@mi.infn.it}
\vspace*{10pt}
\centerline{\footnotesize Bassano Vacchini}
\vspace*{0.015truein}
\centerline{\footnotesize\it Dipartimento di Fisica dell'Universit\`a
  di Milano and INFN,} 
\baselineskip=10pt
\centerline{\footnotesize\it Sezione di Milano,}
\baselineskip=10pt
\centerline{\footnotesize\it 
Via Celoria 16, I-20133 Milan,
Italy}
\baselineskip=10pt
\centerline{\footnotesize\it E-mail: vacchini@mi.infn.it}
\vspace*{0.225truein}
\publisher{(received date)}{(revised date)}

\vspace*{0.21truein} \abstracts{An approach to the description of
  subdynamics inside non-relativistic quantum field theory is
  presented, in which the notions of relevant observable, time scale
  and complete positivity of the time evolution are stressed. A
  scattering theory derivation of the subdynamics of a microsystem
  interacting through collisions with a macrosystem is given, leading
  to a master-equation expressed in terms of the operator-valued
  dynamic structure factor, a two-point correlation function which
  compactly takes the statistical mechanics properties of the
  macrosystem into account. For the case of a free quantum gas the
  dynamic structure factor can be exactly calculated and in the long
  wavelength limit a Fokker-Planck equation for the description of
  quantum dissipation and in particular quantum Brownian motion is
  obtained, where peculiar corrections due to quantum statistics can
  be put into evidence.}{}{}


\vspace*{1pt}\textlineskip      
\section{Introduction}    
\label{intro}
\vspace*{-0.5pt}
\noindent
A subject of major interest in recent research work in
quantum mechanics is the study of time evolutions other than unitary,
allowing for the description of irreversible dynamics. At macroscopic
level the motivation is partly shared with classical physics, lying in
the manifest irreversibility of natural phenomena,\cite{Zeh} and
partly rests on the quest for a clear connection between the extremely
well working quantum mechanical description of physical systems at
microphysical level and our classical perception of reality, hardly
compatible with the superposition 
principle.\cite{Kiefer} At microscopic level the phenomenon which now
attracts most of the attention is decoherence,\cite{Blanchard}
certainly because of its relevance in the understanding of the
appearance of a classical world,\cite{Kiefer} but perhaps even more
because of its fundamental role in answering the question whether
practically useful quantum computers will be feasible in a more or
less distant future;\cite{ZeilingerQC} besides this, non-unitary time
evolutions are essential for the description of quantum dissipation
and approach to equilibrium,\cite{Weiss} issues whose relevance at the
level of applications is increased thanks to the growing ability to
deal experimentally with microphysical probes. The emergence of such
irreversible dynamics is strictly linked to the study of subdynamics,
i.e.\ of the dynamics of a restricted set of degrees of freedom. In the
case of a microsystem this corresponds to the usual procedure in which
one takes the trace over the degrees of freedom of the environment, or
more precisely of the macroscopic system with which the system of
interest is interacting, often leading to a dynamics in which memory
effects can be neglected, describable in terms of a
master-equation. More generally for a system with many degrees of
freedom one considers a subset of relevant observables, suitably
chosen with respect to the quantities that can be effectively measured
on the system, and looks for the subdynamics of this restricted subset
of degrees of freedom, determining the statistical operator
significant for this coarse-grained physical description with
reference to the relevant observables, typically obtaining kinetic
equations.\cite{Roepke,torun99} These effective descriptions
should be meaningful on a coarse-grained time scale over which the
considered observables are suitably slowly varying, typically being
densities of conserved charges.\cite{japan,berlin,slovakia} 

In this paper we shall review some recent work on the formulation of
subdynamics in which the main emphasis lies in the field theoretical
description of the relevant degrees of freedom both for macrosystems
and microsystems, together with a scattering theory approach to the
description of the interaction and a particular attention to the
structural properties of the mappings describing the non-unitary time
evolution, such as complete positivity\cite{Kraus,Hellwig,HolevoNEW}
or a less stringent generalization of it,\cite{berlin,slovakia} 
viewpoints also considered in.\cite{Streater} The
approach has already led to some new results in the treatment of the
subdynamics of a
microsystem, namely in the case of neutron optics,\cite{art2} and most
recently especially in connection with the description of
quantum Brownian motion and of the so-called Rayleigh
gas;\cite{art3,reply,art4,art5,cesena} it is presently under study for the
treatment of subdynamics of relevant observables inside macroscopic
systems,\cite{torun99} as to be discussed later on.
The whole treatment is by now non-relativistic, thus relying on a
second quantization formalism where particle number conservation plays
an important role; some work has however already been done along
similar lines of thought for the generalization to the relativistic
case.\cite{RoepkeQEDI-II} The use of quantum field theory is central
in putting into evidence the interplay between the locality of the
interactions and the confinement pertaining to any real physical
system, expressed through suitable boundary conditions on the fields,
which determine the relevant normal modes. Finiteness of any real
physical system that can be prepared in the laboratory is in fact a
fundamental evidence that can be removed through a thermodynamic
limit, in order to recover more simple and elegant results which may
have general validity, only as a final step, provided finite size
effects are indeed negligible at the chosen level of description. 

The paper is organized as follows: in Sec.~2\ref{conti} the
formalism which leads to a general structure of
master-equation for the description of the subdynamics of a
microsystem is outlined; in Sec.~3\ref{test} its application to the
case of the interaction of a test particle in a quantum gas is
considered; in Sec.~4\ref{cpqbm} the connection to
quantum Brownian motion is discussed; in Sec.~5\ref{end} we comment on
the results and discuss future developments.


\textheight=7.8truein
\setcounter{footnote}{0}
\renewcommand{\thefootnote}{\alph{footnote}}

\section{Field Theoretical Approach to the Derivation of Subdynamics}
\label{conti}
\noindent
We consider a microsystem interacting through collisions with a
macroscopic system, in other words a particle interacting with matter,
both being confined in a finite region which may be taken for
simplicity to be a box, looking for the subdynamics of the
microsystem, essentially referring to,\cite{art1} where a short
derivation of the structure of the master-equation is given, although
a more thorough derivation based on the same physical approximations
can be given and will appear shortly.
 In the absence of external
potentials the Hamiltonian for the particle can be written
\begin{displaymath}
  H_{{\rm \scriptscriptstyle P}}  = \sum_h {E_h} 
  {a^{\scriptscriptstyle \dagger}_{h}}
  a_h, \qquad 
  [a_{h},a^{\scriptscriptstyle  
        \dagger}_{k}]_{\mp}=\delta_{hk}
\end{displaymath}
where $[ A, B]_\mp = A B\mp A B$, $a_h$ and $a^{\scriptscriptstyle
  \dagger}_{k}$ denote annihilation and creation operators for the
particle (obeying either Bose or Fermi statistics) acting in the
Fock-space ${\cal H}_{{\rm \scriptscriptstyle
    P}}=\sum_{n=0}^\infty{}^\oplus {\cal H}_{{\rm \scriptscriptstyle
    P}}^n$ (where ${\cal H}_{{\rm \scriptscriptstyle P}}^n$ is the
symmetrized or antisymmetrized $n$-particle Hilbert space) and the
index $f$ labels a complete set of states $\{ u_f\}$ in ${\cal
  H}_{{\rm \scriptscriptstyle P}}^1$, the normal modes of the
single-particle Hamiltonian with the suitable boundary conditions. The
whole system is then described in the Fock-space ${\cal H}_{{\rm
    \scriptscriptstyle PM}}={\cal H}_{{\rm \scriptscriptstyle
    P}}\otimes {\cal H}_{{\rm \scriptscriptstyle M}}$ by the
Hamiltonian
\begin{displaymath}
  H_{{\rm \scriptscriptstyle PM}}=H_{{\rm \scriptscriptstyle
      P}}+H_{{\rm \scriptscriptstyle
      M}}+V_{{\rm \scriptscriptstyle PM}}, 
\end{displaymath}
where $H_{{\rm \scriptscriptstyle M}}$ describes matter and
satisfies
\begin{displaymath}
  [H_{{\rm \scriptscriptstyle M}},a_f]=0,
\end{displaymath}
while $V_{{\rm \scriptscriptstyle PM}}$ is the interaction
potential. We are interested in the description of a single
microsystem (and therefore the statistics of the microsystem will not
play any role), so that
\begin{displaymath}
  N_{{\rm \scriptscriptstyle P}}=\sum_h 
  {a^{\scriptscriptstyle \dagger}_{h}}
  a_h 
\end{displaymath}
is a conserved quantity, $[V_{{\rm \scriptscriptstyle
    PM}},N_{{\rm \scriptscriptstyle P}}]=0$ and as a consequence
$[H_{{\rm \scriptscriptstyle 
    PM}},N_{{\rm \scriptscriptstyle P}}]=0$. We therefore only
describe scattering without absorption or creation phenomena,
according to the non-relativistic treatment. Since we are considering
a single particle we take for the statistical operator describing the
whole system at the initial time the following uncorrelated expression:  
        \begin{equation}
        {\rho}_{{\rm \scriptscriptstyle PM}}=
        \sum_{gf}{}
        {a^{\scriptscriptstyle \dagger}_{g}}
        \varrho_{{\rm \scriptscriptstyle M}} {a_{{f}}}
        {\varrho}_{gf}    ,
        \label{25}
        \end{equation}
where $\varrho_{{\rm \scriptscriptstyle M}}$ is a statistical operator
in 
${\cal
  H}_{{\rm \scriptscriptstyle PM}}^0\equiv {\cal
  H}_{{\rm \scriptscriptstyle P}}^0\otimes {\cal
  H}_{{\rm \scriptscriptstyle M}},$ the subspace of ${\cal
  H}_{{\rm \scriptscriptstyle PM}}$ in which
$N_{{\rm \scriptscriptstyle P}}=0$, describing the macroscopic
system alone, so that
        \begin{equation}
        \label{zero}
        {a_{{f}}}\varrho_{{\rm \scriptscriptstyle M}}=
        \varrho_{{\rm \scriptscriptstyle M}}
        {a^{\scriptscriptstyle \dagger}_{f}}=0 \quad
        \forall f
        .
        \end{equation}
In terms of the conserved charge $Q=N_{{\rm \scriptscriptstyle
    P}}$ eq.~(\ref{zero}) means that $\varrho_{{\rm \scriptscriptstyle
    M}}$ has charge 
zero, $ Q {\varrho}_{\rm  \scriptscriptstyle M}=0$, i.e.\ the
microsystem is not part of 
the macrosystem, while (\ref{25}) means that
$\rho_{\rm \scriptscriptstyle PM}$ describes the system
perturbed by a single microsystem, i.e.:
\begin{displaymath}
  Q\rho_{\rm \scriptscriptstyle PM}=\rho_{\rm \scriptscriptstyle PM}.
\end{displaymath}
We will assume that the macrosystem is not appreciably perturbed by
the presence of the microsystem, so that its dynamics is given by
\begin{displaymath}
  {
    d{\varrho_{{\rm \scriptscriptstyle M}}}
    \over
    dt
    }  =
  - {i \over \hbar} [{H}_{\rm  \scriptscriptstyle M},{\varrho_{{\rm
        \scriptscriptstyle M}}}]  . 
\end{displaymath}
The coefficients ${{\varrho}}_{gf}$ in (\ref{25}) build a positive,
trace one matrix, to be seen as the representative of a statistical
operator ${{\hat \varrho}}$ in ${\cal H}_{{\rm \scriptscriptstyle
    P}}^1$ spanned by the states $\{ u_f\}$ according to $
{{\varrho}}_{gf}= \langle u_g | {{\hat \varrho}} | u_f \rangle $, so
that $\rho_{\rm \scriptscriptstyle PM}$ is indeed a statistical
operator. According to the general purpose we are only interested in
the subdynamics of a subset of slowly varying observables, generally
given by linear operators in ${\cal H}_{{\rm \scriptscriptstyle
    PM}}$, and not in a dynamics to be considered reliable for any
observable of the system. In this specific case the relevant degree
of freedom is the particle, whose subdynamics we are looking for, so
that we restrict to operators of the form
\begin{equation}
        {A}
        = \sum_{h k}
        {a^{\scriptscriptstyle \dagger}_{h}}
        A_{hk}
        {a_{k}}
        = \sum_{h k}
        {a^{\scriptscriptstyle \dagger}_{h}}
        \langle
        u_h
        \vert
        {\hat {\sf A}}
        \vert
        u_k
        \rangle
        {a_{k}},
        \label{23}
\end{equation}
where ${\hat {\sf A}}$ can generally be an operator in ${\cal
  H}_{{\rm \scriptscriptstyle PM}}^1={\cal
  H}_{{\rm \scriptscriptstyle P}}^1\otimes {\cal
  H}_{{\rm \scriptscriptstyle M}}$ or equivalently $A_{hk}$ can be
operator-valued in ${\cal H}_{{\rm \scriptscriptstyle M}}$. In
order to determine the dynamics of the microsystem we consider the
following simple reduction formula from ${\cal
  H}_{{\rm \scriptscriptstyle PM}}$ to ${\cal
  H}_{{\rm \scriptscriptstyle P}}^1$ for the expectation value of
observables of the form (\ref{23}) in the state (\ref{25}): 
\begin{equation}
        \textrm{Tr}_{{\cal H}_{\rm \scriptscriptstyle PM}}
        \left(
        {{A}{\rho}_{\rm \scriptscriptstyle PM}}
        \right)
        = 
        \sum_{h k}
        {\varrho}_{kh} {\rm Tr}_{{\cal
            H}_{{\rm \scriptscriptstyle M}}} (A_{hk} \varrho_{\rm
          \scriptscriptstyle M}) = \sum_{h k}{\varrho}_{kh} \bar{A}_{hk}
         =
        {\rm Tr}_{{\cal
            H}^1_{{\rm \scriptscriptstyle P}}} (\hat{\varrho}
        \hat{{\bar{\sf A}}}) 
        ,
        \label{reduction}
\end{equation}
where
\begin{displaymath}
  \bar{A}_{hk}=\langle u_h|{\rm Tr}_{{\cal
            H}_{{\rm \scriptscriptstyle M}}} (A \varrho_{\rm
          \scriptscriptstyle M}) |u_k \rangle=\langle
        u_h|\hat{{\bar{\sf A}}}|u_k \rangle . 
\end{displaymath}
Let us note that even if $A_{hk}$ is initially a c-number, it
becomes operator-valued in ${\cal H}_{{\rm \scriptscriptstyle M}}$ due
to the time evolution.

In order to obtain the subdynamics of the
particle, given by the time dependence of the coefficients
$\varrho_{gf}$, we are led to consider in particular the operator
\begin{displaymath}
  A=a^{\scriptscriptstyle \dagger}_{f} a_g ,
\end{displaymath}
so that ${\hat {\sf A}}$ is given by the rank one operator
$|u_f\rangle\langle u_g |$ and one has, according to (\ref{reduction})
\begin{displaymath}
\textrm{Tr}_{{\cal H}_{\rm \scriptscriptstyle PM}}
        \left(
        {{A}{\rho}_{\rm \scriptscriptstyle PM}}
        \right)
        =  \varrho_{gf} .
\end{displaymath}
The microsystem represents here the selected degree of freedom, with a
characteristic variation time $\tau$ which is much longer than the
relaxation time of the macrosystem, which is a microphysical time
$\tau_0$, typically of the order the duration of a collision. The slow
variability will naturally depend on the physical features of the
normal modes $\{ u_f\}$ and of the interaction $V_{{\rm
    \scriptscriptstyle PM}}$. We determine the generator of
the time evolution of the statistical operator for the microsystem,
which according to irreversibility will be generally given by a
semigroup,\cite{Alicki,HolevoNEW} on a time scale $\tau$ much longer
than the correlation time for the macrosystem, approximating $ { d
  {\varrho}_{gf}(t) / dt } $ by:
        \begin{equation}
        {
        \Delta {\varrho}_{gf}(t)
        \over
        \tau
        }
        =
        {1\over \tau}
        \left[
        {\varrho}_{gf}(t+\tau) -
        {\varrho}_{gf}(t)
        \right]
        =
        {1\over \tau}
        \left[
        \textrm{Tr}_{{\cal H}_{\rm \scriptscriptstyle PM}}
        \left(
        {a^{\scriptscriptstyle \dagger}_{f}} {a_{g}}
        e^{-{{
        i
        \over
         \hbar
        }}H\tau}
        \varrho (t)
        e^{{{
        i
        \over
         \hbar
        }}H\tau}
        \right)
        -
        {\varrho}_{gf}(t)
        \right]
        .
        \label{new}
        \end{equation}
We then exploit the cyclic invariance of the trace, working in
Heisenberg picture and shifting the action of the temporal evolution
operator on the simple operator expression ${a^{\scriptscriptstyle
    \dagger}_{f}} {a_{g}}$, thus concentrating on the observables and
considerably simplifying the calculation, without introducing
restrictive assumptions on the structure of $\varrho_{{\rm
    \scriptscriptstyle M}}$ or 
of the interaction.  We have to study the expression $e^{+{i\over
    \hbar}{H}_{\rm \scriptscriptstyle PM}t}
{a}_h^{\scriptscriptstyle\dagger} {a}_k e^{-{i\over 
    \hbar}{H}_{\rm \scriptscriptstyle PM}t}$, relying on the slow
variability of the considered 
observable, corresponding to the quasi-diagonality in the indexes
$h,k$.  To proceed further we introduce the following superoperators
(the prime recalling the Heisenberg picture)
        \begin{equation}
        \label{super}
        {\cal H}'={i \over \hbar} [H_{{\rm \scriptscriptstyle
            PM}},\cdot],  \quad 
        {\cal H}'_0={i \over \hbar} [H_{{\rm \scriptscriptstyle P}} +
        H_{{\rm \scriptscriptstyle M}},\cdot], 
        \quad
        {\cal V}'={i \over \hbar} [V_{{\rm \scriptscriptstyle
            PM}},\cdot]       , 
        \end{equation}
acting
on the algebra generated by creation and annihilation
operators.
Note that the operators
$
 ({a^{\scriptscriptstyle \dagger}_{h_1}})^{n_1}
({a^{\scriptscriptstyle \dagger}_{h_2}})^{n_2}
$
$\ldots$
$
({a^{\scriptscriptstyle \dagger}_{h_r}})^{n_r}
$
$
({a_{k_1}})^{m_1}
({a_{k_2}})^{m_2}
$
$\ldots$
$
(a_{k_s})^{m_s}  
$
are {\em eigenvectors\/} of the superoperator ${\cal H}'_0$ with
eigenvalues
$
{i \over \hbar}
\left(
{\sum_{i=1}^{r}n_i E_{h_i} - \sum_{i=1}^{s}m_i E_{k_i}}
\right)
$, in particular:
        \[
        {\cal H}'_0 {a_{{h}}} = -{i \over \hbar} {E_h} {a_{{h}}} \qquad
        {\cal H}'_0 {a^{\scriptscriptstyle \dagger}_{h}} =
        +{i \over \hbar} {E_h} {a^{\scriptscriptstyle \dagger}_{h}}  .
        \]
In order to calculate (\ref{new}) we 
set ${\cal U}'(t)=e^{{\cal H}'t}$
and evaluate
${\cal U}'(t)\left(
{{a^{\scriptscriptstyle \dagger}_{h}}{a_{k}}}
\right)
$
by means of the following integral representation:        \[  
        {{\cal U}'(t){a_{k}}}  
        =  
        {\int_{-i\infty+\eta}^{+i\infty +  \eta}}{  
        dz  
        \over  
            2\pi i  
        }       \,   e^{z t}  
        {  
        {{  
        \left(  
        {{ z - {\cal H}'}}  
        \right)  
        }^{-1}}  
        {a_{k}}}  
        ,  
        \qquad  
        {\cal U}'(t)  
        \left(  
        {{a^{\scriptscriptstyle \dagger}_{h}}{a_{k}}}  
        \right)  
        =  
        \left(  
        {\cal U}'(t){a^{\scriptscriptstyle \dagger}_{h}}  
        \right)  
               \left(  
        {{\cal U}'(t){a_{k}}}  
               \right)  
        .  
        \]  

For the mappings defined in (\ref{super}) identities hold that are
reminiscent of the usual ones in scattering theory:  
        \begin{equation}  
        {{  
        \left(  
        {{ z - {\cal H}'}}  
        \right)  
        }^{-1}}  
        = {{  
        \left(  
        {{ z - {\cal H}'_0}}  
        \right)  
        }^{-1}}  
        \left[{1+{\cal V}'{{  
        {\left( 
        {{ z - {\cal H}'}}  
        \right)} 
        }^{-1}}}\right] =  
        {\left[{1+{{ 
        {\left( 
        {{ z - {\cal H}'}}  
        \right)} 
        }^{-1}}{\cal V}'}\right]}{{ 
        \left(  
        {{ z - {\cal H}'_0}}  
        \right)  
        }^{-1}}  .  
        \label{a9}  
        \end{equation}  
In particular we can introduce the superoperator  
${{\cal T}(z)}$  
        \begin{equation}  
        {\cal T}(z)  
        \equiv  
        {\cal V}' + {\cal V}'{{  
        \left(  
        {{ z - {\cal H}'}}  
        \right)  
        }^{-1}}{\cal V}',  
        \label{a10}  
        \end{equation}  
satisfying  
\begin{equation}
        {{  
        \left(  
        {{ z - {\cal H}'}}  
        \right)  
        }^{-1}}={{  
        \left(  
        {{ z - {\cal H}'_0}}  
        \right)  
        }^{-1}} +{{  
        \left( 
        {{ z - {\cal H}'_0}}  
        \right) 
        }^{-1}}  
        {\cal T}(z){{  
        \left(  
        {{ z - {\cal H}'_0}}  
        \right)  
        }^{-1}}  \label{a11}    
\end{equation}
and  
        \begin{equation}  
        {\cal T}(z)=  
        {\cal V}' + {\cal V}'{{  
        \left(  
        {{ z - {\cal H}'_0}}  
        \right)  
        }^{-1}}{\cal T}(z),  
        \label{a14}  
        \end{equation}  
corresponding to the Lippman-Schwinger equation for the T-matrix.
Exploiting $[H_{{\rm \scriptscriptstyle PM}},N_{{\rm
    \scriptscriptstyle P}}]=0$  
the restriction to ${{{\cal H}^1_{{\rm \scriptscriptstyle PM}}}}$  of the  
operator ${{\cal T}(z)}[{a_{k}}]$ has the
simple general form:  
        \begin{equation}  
        \label{a13}  
        i\hbar  
        {{{\cal T}(z)}[{a_{k}}]}_{|{{{\cal H}^1_{{\rm
                  \scriptscriptstyle PM}}}}} 
        =\sum_f
        T{}_{f}^{k}
        \left(  
         i\hbar  z  
        \right)  
        {a_{{f}}} ,
        \end{equation}  
and similarly, taking the adjoint
        \[
        {}-i\hbar
        {{{\cal T}(z^*)}
        [{a^{\scriptscriptstyle\dagger}_{h}}]
        }_{|{{{\cal H}^0_{{\rm \scriptscriptstyle PM}}}}}
        =\sum_f
        {[T{}_{f}^{h}\left(
        i\hbar  z
        \right)]}^{\scriptscriptstyle\dagger}
        a_f^{\scriptscriptstyle\dagger}
        \equiv
        \sum_f
        {T{}_{f}^{h}{}^{\scriptscriptstyle\dagger}\left(
        i\hbar  z
        \right)}
        a_f^{\scriptscriptstyle\dagger},
        \]
where $  
T{}_{f}^{k}
\left(  
   z  
\right)  
$ is an operator in the subspace  
${{{\cal H}^0_{{\rm \scriptscriptstyle PM}}}}$.  
This restriction is the only part of interest to us, since we are  
considering a single microsystem.  
One can also express $
T{}_{f}^{k}
\left(
   z
\right)
$ in terms of ${{\cal T}(z)}$ as:
        \begin{eqnarray}
        \label{212a}
        {i\hbar {\cal T}(z)
        [a_k] a_h^{\scriptscriptstyle\dagger}
        }_{|{{{\cal H}^0_{{\rm \scriptscriptstyle PM}}}}}
        \nonumber
        &=&
        T{}_h^k (i\hbar z)
        \\
        -i\hbar
        a_h
        {\cal T}(z) [a_k^{\scriptscriptstyle\dagger}
        ]_{|{{{\cal H}^0_{{\rm \scriptscriptstyle PM}}}}}
        &=&
        T{}_h^k{}^{\scriptscriptstyle\dagger}
        (i\hbar z^*)
        .
        \end{eqnarray}

Denoting by ${\vert \lambda \rangle}\equiv
\vert 0 \rangle
\otimes{\vert \lambda \rangle}$
the basis of eigenstates of
$H_{{\rm \scriptscriptstyle M}}$ spanning ${{{\cal H}^0_{{\rm
        \scriptscriptstyle PM}}}}$, 
        $
        H_{{\rm \scriptscriptstyle M}}{\vert \lambda \rangle}=
        E_\lambda
        {\vert \lambda \rangle},
        $
and
exploiting
(\ref{a13})
we
obtain
the following explicit representation of $
{{\cal U}'(t)a_{k}}_{|{\cal H}^1_{{{\rm \scriptscriptstyle
                PM}}}}
$ as a mapping of
${{{\cal H}^1_{\rm \scriptscriptstyle PM}}}$
into ${{{\cal H}^0_{{\rm \scriptscriptstyle PM}}}}$:
        \begin{eqnarray}
        \label{bad}  
        &&
        \!\!\!\!\!\!\!\!\!
        {{\cal U}'(t)a_{k}}_{|{\cal H}^1_{{{\rm \scriptscriptstyle
                PM}}}}
        =
        {\int_{-i\infty+\eta}^{+i\infty +  \eta}}{
        dz
        \over
            2\pi i
        }       \,   e^{z t}
        \left[
        {
        ( z-{\cal H}'_0)^{-1}
        }
        +
        {
        ( z-{\cal H}'_0)^{-1}
        }
        {\cal T}(z)
        {
         ( z-{\cal H}'_0)^{-1}
        }
        \right]
        {a_{k}}_{|{\cal H}^1_{{{\rm \scriptscriptstyle PM}}}}
         \nonumber\\
        &&=
        {\int_{-i\infty+\eta}^{+i\infty +  \eta}}{
        dz
        \over
            2\pi i
        }       \,   e^{z t}
        {
        1
        \over
         z + \frac{i}{\hbar} E_k
        }
        \left[
        a_k
        +
        {
        ( z-{\cal H}'_0)^{-1}
        }
        {
        1
        \over
         i\hbar
        }
        \sum_f T{}_f^k(i\hbar z)
        a_f
        \right]
         \nonumber\\
        &&=
        e^{- \frac{i}{\hbar}E_k t} a_k
        +
        \frac{1}{i\hbar}
        \sum_{\lambda{\lambda'} \atop f}
        {\int_{-i\infty+\eta}^{+i\infty +  \eta}}{
        dz
        \over
            2\pi i
        }       \,   e^{z t}
        {
        \vert {\lambda'} \rangle \langle {\lambda'} \vert
        T{}_f^k(i\hbar z)
        \vert \lambda \rangle
        \langle \lambda \vert
        \over
        \left(
        {
        z + \frac{i}{\hbar} E_k
        }
        \right)
        \left(
        z + \frac{i}{\hbar}     (E_f + E_\lambda -
        E_{\lambda'})
        \right)
        }
        a_f  .
         \nonumber\\
        \end{eqnarray}
The operator ${\cal T}(z)$ has poles on
the imaginary axis for $z=(i/ \hbar)(e_\alpha-e_\beta)$,
$e_\alpha$ being the eigenvalues of $H_{{\rm \scriptscriptstyle PM}}$. In the
calculation we assume that the
function ${\cal T}(z)$ for Re$\>z$ positive and much bigger than the
typical spacing between the   
poles is smooth enough, so that the only relevant contribution  
stems from
the singularities of $({ z - {\cal H}'_0})^{-1}$: this smoothness
property is linked to the fact that the set of poles of $({  
z - {\cal H}'})^{-1}$ goes over to a continuum if the confinement is
removed yielding an analytic function with a cut along the imaginary  
axis, that can be continued across the cut without singularities   
if no absorption of the  microsystem occurs. The T-matrix will
therefore be taken to depend very smoothly on energy. Evaluating the
integral (\ref{bad}) becomes
        \begin{eqnarray*}
        {{\cal U}'(t)a_{k}}_{|{\cal H}^1_{{{\rm \scriptscriptstyle PM}}}}
        =&+&
        e^{- \frac{i}{\hbar}E_k t} a_k
        \\
        &+&
        \sum_{\lambda {\lambda'} \atop f}
        \vert {\lambda'} \rangle
        \left[
          e^{- \frac{i}{\hbar}E_k t}
        {
        \langle {\lambda'} \vert
        T{}_f^k(E_k)
        \vert \lambda \rangle
        \over
        E_k + E_{\lambda'} - E_f - E_\lambda
        }
        \right.
        \\
        &&
        \left.
        \hphantom{spost}  
        {}+
        e^{-\frac{i}{\hbar}(E_f - E_\lambda -E_{\lambda'})t}
        {
        \langle {\lambda'} \vert
        T{}_f^k
        \left(
        E_f + E_\lambda -E_{\lambda'}
        \right)
        \vert \lambda \rangle
        \over
        E_f + E_\lambda -E_{\lambda'}
        -
        E_k
        }
        \right]
        \langle \lambda \vert
        a_f,
        \end{eqnarray*}
and similarly for the adjoint mapping. 
On a time scale $t$, much longer than the collision time
$\tau_0$, but still much shorter than the typical variation time
inside the reduced description $\tau$ 
($\tau_0 \ll t \ll\tau$), considering suitable
slow variables, so that
        \begin{displaymath}
        {
        \left |
        E_h - E_k  
        \right |
        \over
        \hbar
        } \ll
        {  
        1
        \over  
             \tau_0  
        } ,
        \qquad
        {
        \left |
        E_h+E_{\lambda'} - E_f - E_\lambda
        \right |
        \over
        \hbar
        } \ll
        {
        1
        \over
             \tau_0
        }              ,
        \end{displaymath}
one obtains the following expression for  the generator of
the time evolution in Heisenberg picture, denoted by ${\cal L}'$ 
        \begin{eqnarray}
        \label{serve}
        &&
        \!\!\!\!\!\!\!\!{{\cal U}'(t)
        \left(
          a_h^{\scriptscriptstyle\dagger}
        a_{k}
        \right)}_{|{\cal H}^1_{{{\rm \scriptscriptstyle PM}}}}
        =
        a_h^{\scriptscriptstyle\dagger}
        a_{k}   +t {\cal L}'
        \left(
        a_h^{\scriptscriptstyle\dagger}
        a_{k}
        \right)
        \nonumber
        \\
        &&=
        a_h^{\scriptscriptstyle\dagger}
        a_{k}
        +\frac{i}{\hbar}
        t (E_h -E_k)
        a_h^{\scriptscriptstyle\dagger}
        a_{k}
        \nonumber
        \\
        &&
        \hphantom{=}{}
        -\frac{i}{\hbar}
        t \sum_f a_h^{\scriptscriptstyle\dagger}
        T{}_f^k(E_k+i\varepsilon)
        a_f
        +\frac{i}{\hbar} t \sum_g
        a_g^{\scriptscriptstyle\dagger}
        T{}_g^h{}^{\scriptscriptstyle\dagger}(E_h+i\varepsilon)
        a_k
        \nonumber
        \\
        &&
        \hphantom{=}
        {}+
        2\frac{\varepsilon}{\hbar}
        t \sum_{\lambda\alpha\alpha'  \atop gf}
        a^{\scriptscriptstyle\dagger}_g
        \vert\alpha\rangle
        {
        \langle\alpha\vert
        T{}_g^h{}^{\scriptscriptstyle\dagger}(E_h+i\varepsilon)
        \vert\alpha'\rangle
        \over
        E_g + E_{\alpha} -E_h-E_{\alpha'}-i\varepsilon
        }
        {
        \langle\alpha'\vert
        T{}_f^k(E_k+i\varepsilon)
        \vert \lambda \rangle
        \over
        E_f + E_{\lambda} -E_k-E_{\alpha'}+i\varepsilon
        }
        \langle \lambda \vert
        a_f,
        \nonumber
      \\
      \end{eqnarray}
where $\varepsilon$ is a positive quantity, which can tend to zero
after the thermodynamic limit has been taken. In view of (\ref{serve})
let us define the operators
        \begin{eqnarray*}
        { T}^{[1]}
        &\!\!=\!\!&
        \sum_{gr}
        { a}^{\scriptscriptstyle \dagger}_{r}
        T{}_g^r(E_r+i\varepsilon)
        { a}_{g}
        \\
        {T}^{[1]}{}^{\dagger}
        &\!\!=\!\!&
        \sum_{gr}
        { a}^{\scriptscriptstyle \dagger}_{r}
        T{}_r^g{}^{\scriptscriptstyle\dagger}(E_g+i\varepsilon)
        { a}_{g}
        \nonumber \\
        R{}_{k\lambda}^{[1]}{}^{\hphantom{{\dagger}}}
        &\!\!=\!\!&
        \sum_f R_{k\lambda f}a_f
        =
        \sum_f
        \left[
        \sqrt{2\varepsilon}
        \langle \lambda \vert
        T{}_f^k(E_k+i\varepsilon)
        {
        \left( E_f + H_{\rm \scriptscriptstyle M}
          -E_k-E_{\lambda}+i\varepsilon  \right)^{-1} 
        }
        \right]
        a_f
        \nonumber \\
        R{}_{h\lambda}^{[1]}{}^{\dagger}
        &\!\!=\!\!&
        \sum_g   a^{\scriptscriptstyle\dagger}_g
        R_{h\lambda g}^{\scriptscriptstyle\dagger}
        =
        \sum_g
        a^{\scriptscriptstyle\dagger}_g
        \left[
        \sqrt{2\varepsilon}
        {
        \left( E_g + H_{\rm \scriptscriptstyle M}
          -E_h-E_{\lambda}-i\varepsilon \right)^{-1} 
        }
        T{}_g^h{}^{\scriptscriptstyle\dagger}(E_h+i\varepsilon)
        \vert \lambda \rangle
        \right],
        \nonumber
        \end{eqnarray*}
so that we can write
        \[
        {\cal L}'
        \left(  
        {{{ a}^{\scriptscriptstyle \dagger}_{h}}{{ a}_{k}}}
        \right)  
        =
        {i\over\hbar}  
        \left[  
        { H}_0,
        { a}^{\scriptscriptstyle \dagger}_{h}
        { a}_{k}
        \right]  
        + {i\over \hbar}
        \left[  
        {T}^{[1]}{}^{\dagger} ,
        { a}^{\scriptscriptstyle\dagger}_h
        \right]  
        { a}_k
        + {i\over \hbar}
        { a}^{\scriptscriptstyle\dagger}_h
        \left[  
        {T}^{[1]}, { a}_k
        \right]  
        +  
        {1\over\hbar} \sum_\lambda  
        { R}^{[1]}_{h \lambda}{}^{\dagger}
        { R}^{[1]}_{k \lambda}.
        \]
Introducing the following one-particle operators
        \begin{displaymath}
        \begin{array}{rcccccl}
        { V}^{[1]}
        &=&
        \sum_{gr}  
        { a}^{\scriptscriptstyle \dagger}_{r}
        { V}{}_{rg}
        { a}_{g}
        &=&
        \frac 12
        \left[
        {T}^{[1]} +
        {T}^{[1]}{}^{\dagger}
        \right]
        \\
        &&\hphantom{tutto}&&
        \\
        {\Gamma}^{[1]}
        &=&
        \sum_{gr}
        { a}^{\scriptscriptstyle \dagger}_{r}
        \Gamma_{rg}
        { a}_{g}
        &=&
        \frac i2
        \left[
        {T}^{[1]} -
        {T}^{[1]}{}^{\dagger}
        \right]
        \end{array}
        \end{displaymath}
so that
        \[
        {T}^{[1]}
        =
        {V}^{[1]} -i
        {\Gamma}^{[1]} ,
        \qquad
        {V}^{[1]}={V}^{[1]}{}^{\dagger},
        \qquad
        {\Gamma}^{[1]}={\Gamma}^{[1]}{}^{\dagger}
        \]
the generator of the time evolution may be written
        \begin{equation}
        \label{generatore}
        {\cal L}'
        \left(  
        {{{ a}^{\scriptscriptstyle \dagger}_{h}}{{ a}_{k}}}
        \right)  
        =
        {i\over\hbar}  
        \left[  
        { H}_0 + V^{[1]},
        { a}^{\scriptscriptstyle \dagger}_{h}
        { a}_{k}
        \right]  
        - {1\over \hbar}  
        \left\{
        \left[  
        {\Gamma}^{[1]} , { a}^{\scriptscriptstyle\dagger}_h
        \right]  
        { a}_k
        -  
        { a}^{\scriptscriptstyle\dagger}_h
        \left[  
        {\Gamma}^{[1]}, { a}_k
        \right]  
        \right\}
        +  
        {1\over\hbar} \sum_\lambda  
        { R}^{[1]}_{h \lambda}{}^{\dagger}
        { R}^{[1]}_{k \lambda}.
        \end{equation}
Let us observe that
${ V}_{rg}$  and ${ \Gamma}_{rg}$ are not c-number
coefficients, but operators acting in the Fock-space for the  
macrosystem ${{{\cal H}_{{\rm \scriptscriptstyle M}}}}$: they are
connected respectively  
to the self-adjoint and anti-self-adjoint part of  
what can be considered as an operator valued T-matrix.  
The last contribution displays a {\em bilinear  
structure\/} typical of the generators of completely positive time
evolutions,\cite{slovakia} directly connected to irreversibility, as we
shall see in Sec.~2.1\ref{pnc}. 

\subsection{Particle Number Conservation and Complete Positivity}
\label{pnc}
\noindent
It is worth mentioning that within the approximations exploited in its
derivation (\ref{generatore}) accounts for particle number
conservation, that is to say ${\cal L}'(N_{{\rm \scriptscriptstyle
    P}})=0$ and therefore ${\cal U}'(t)(N_{{\rm \scriptscriptstyle
    P}})=N_{{\rm \scriptscriptstyle P}}$. Due to
        \[
        \sum_h
        \left[
        {\Gamma}^{[1]} , { a}^{\scriptscriptstyle\dagger}_h
        \right]  
        { a}_h
        =
        {\Gamma}^{[1]}     ,
        \qquad
        \sum_h a^{\scriptscriptstyle\dagger}_h
        \left[  
        {\Gamma}^{[1]}, { a}_h
        \right]  
        =
        -        {\Gamma}^{[1]}
        \]
we have
        \[
        {\cal L}'(N_{{\rm \scriptscriptstyle P}})= - \frac{2}{\hbar}
        {\Gamma}^{[1]} + \frac{1}{\hbar} \sum_{h\lambda}
        R_{h\lambda}^{[1]}{}^{\dagger}
        R_{h\lambda}^{[1]}                           ,
        \]
and therefore particle number conservation amounts to
\begin{equation}
\label{N}
        {\Gamma}^{[1]}  =
        \frac{1}{2} \sum_{h\lambda}
        R_{h\lambda}^{[1]}{}^{\dagger}
        R_{h\lambda}^{[1]}.
\end{equation}
 
The structure of (\ref{generatore})  is moreover such that ${\cal
U}'(t)$ satisfies a property analogous to, but weaker than complete
positivity.
We first briefly recall the definition of complete
positivity.\cite{Kraus,Hellwig,HolevoNEW} 
Consider a system described in a Hilbert space $H$ and the set $B (H)$ 
of bounded linear operators on $H$, containing the observables. A
mapping ${\cal M}'$ defined on this set,
\begin{displaymath}
  {\cal M}':B (H)\rightarrow B (H),
\end{displaymath}
e. g. a mapping giving the dynamics in Heisenberg picture, is said to be
completely positive provided it satisfies the inequality
\begin{equation}
\label{cpvera}
        \sum_{i,j=1}^n
        \langle\psi_i\vert
        {\cal M}'
        \left(
        {\hat{B}}{}_i^{\scriptscriptstyle\dagger}
        {\hat {B}}{}_j
        \right)
        \vert\psi_j\rangle
        \geq 0
        \qquad
        \forall n\in{\bf N}, \quad
        \forall
        \left \{
        \psi_i
        \right \}
        \in { H}
        , \quad
        \forall
        \{
        {\hat B}_i
        \}
        \in { B}({ H}).
        \end{equation}
For $n=1$ the usual notion of positivity is obtained,
the condition $n\in\textbf{N}$ implying that complete positivity is
in general actually a stronger requirement. It is worthwhile observing 
that unitary evolutions are not only positive, but also completely positive.
Eq.~(\ref{generatore}) satisfies a weaker version of complete
positivity\cite{berlin,slovakia} 
in that an inequality like (\ref{cpvera}) only holds for a restricted
subset of observables, bilinear in the  creation and
annihilation  operators $a$ and $a^{\scriptscriptstyle\dagger}$, that
is to say
\begin{equation}
        \sum_{i,j=1}^n
        \langle\psi_i\vert
        {\cal U}'(t)
        \left[
        \sum_{hk}
        a^{\scriptscriptstyle\dagger}_h
        \langle u_h\vert {\hat{\sf
        B}}{}_i^{\scriptscriptstyle\dagger}
        {\hat {\sf B}}{}_j \vert u_k \rangle    a_k
        \right]
        \vert\psi_j\rangle
        \geq 0
        \qquad        \forall n\in{\bf N}
        \label{cp}
\end{equation}
with 
$\left \{ \psi_i \right \} $ vectors in Fock-space and $ \{ {\hat {\sf
    B}}_i \} $ operators in the one-particle Hilbert space ${{{\cal
      H}^1_{{\rm \scriptscriptstyle P}}}}$.
Setting
        \[
        \left[  
        {\Gamma}^{[1]}, { a}_k
        \right]  
        = - F_k      ,
        \qquad
        \left[  
        {\Gamma}^{[1]} , { a}^{\scriptscriptstyle\dagger}_h
        \right]  
        = F_h^{\scriptscriptstyle\dagger},
        \qquad
        H_0 + V^{[1]}_{\hphantom{0}} =C
        \]
we have in fact, at first order in $t$
        \begin{eqnarray}
          \label{brutta}
          \nonumber
        &&
        \!\!\!\!\!\!\!\!\!\!\!\!\!\!\!\!\!\!\!\!\!\!\!\!\!\!
        \sum_{i,j=1}^n
        \langle\psi_i\vert
        {\cal U}'(t)
        \left[
        \sum_{hk}
        a^{\scriptscriptstyle\dagger}_h
        \langle  u_h
        \vert {\hat{\sf B}}{}_i^{\scriptscriptstyle\dagger}
        {\hat {\sf B}}{}_j \vert u_k \rangle a_k
        \right]
        \vert\psi_j\rangle
        =
        \\
        \nonumber &=&
        \sum_{i,j=1}^n  \sum_{hk}
        \langle u_h\vert {\hat{\sf
        B}}{}_i^{\scriptscriptstyle\dagger}
        {\hat {\sf B}}{}_j \vert u_k \rangle
        \langle\psi_i\vert
        a^{\scriptscriptstyle\dagger}_h a_k + t
        {\cal L}' (a^{\scriptscriptstyle\dagger}_h a_k)
        \vert\psi_j\rangle
        \\
        \nonumber &=&
        \sum_{i,j=1}^n  \sum_{hk}
        \langle u_h\vert {\hat{\sf
        B}}{}_i^{\scriptscriptstyle\dagger}
        {\hat {\sf B}}{}_j \vert u_k \rangle
        \langle\psi_i\vert
        \left[
        a_h+t\frac{i}{\hbar}
        \left[
        C,a_h
        \right]
        -\frac{t}{\hbar} F_h
        \right]^{\dagger}
        \\
        \nonumber &&
        \hphantom{\sum_{i,j=1}^n\sum_{hk}\>}
        \times
        \left[
        a_k+t\frac{i}{\hbar}
        \left[
        C,a_k
        \right]
        -\frac{t}{\hbar} F_k
        \right]
        + \frac{t}{\hbar} \sum_\lambda
        { R}^{[1]}_{h \lambda}{}^{\scriptscriptstyle\dagger}
        { R}^{[1]}_{k \lambda}
        \vert\psi_j\rangle
        \\
        \nonumber &=&
        \sum_g
        \left(
        \sum_{i=1}^n
        \sum_h
        \langle u_h\vert {\hat{\sf
        B}}{}_i^{\scriptscriptstyle\dagger}
        \vert u_g \rangle
        \langle\psi_i\vert
        \left[
        a_h+t\frac{i}{\hbar}
        \left[
        C,a_h
        \right]
        -\frac{t}{\hbar} F_h
        \right]^{\dagger}
        \right)
        \\
        \nonumber &&
        \hphantom{
        ==
        }
        \times
        \left(
        \sum_{j=1}^n
        \sum_k        
        \langle u_g\vert
        {\hat {\sf B}}{}_j
        \vert u_k \rangle
        \left[
        a_k+t\frac{i}{\hbar}
        \left[
        C,a_k
        \right]
        -\frac{t}{\hbar} F_k
        \right]
        \vert\psi_j\rangle
        \right)
        \\
        \nonumber &&
        + \frac{t}{\hbar} \sum_{g\lambda}
        \left(
        \sum_{i=1}^n
        \sum_h
        \langle u_h\vert {\hat{\sf
        B}}{}_i^{\scriptscriptstyle\dagger}
        \vert u_g \rangle
        \langle\psi_i\vert
        { R}^{[1]}_{h \lambda}{}^{\dagger}
        \right)
        \left(
        \sum_{j=1}^n
        \sum_k
        \langle u_g\vert
        {\hat {\sf B}}{}_j
        \vert u_k \rangle
        { R}^{[1]}_{k \lambda}
        \vert\psi_j\rangle
        \right)
        \\
        \nonumber &=&
        \sum_g
        \Biggl\|
        \sum_{j=1}^n   
        \sum_k
        \langle u_g\vert
        {\hat {\sf B}}{}_j
        \vert u_k \rangle
        \left[
        a_k+t\frac{i}{\hbar}
        \left[
        C,a_k
        \right]
        -\frac{t}{\hbar} F_k
        \right]
        \vert\psi_j\rangle
        \Biggr\|^2_{{\cal H}_{{\rm \scriptscriptstyle PM}}}
        \\
        &&
        {}+\frac{t}{\hbar} \sum_{g\lambda}
        \Biggl\|
        \sum_{j=1}^n
        \sum_k
        \langle u_g\vert
        {\hat {\sf B}}{}_j
        \vert u_k \rangle
        { R}^{[1]}_{k \lambda}
        \vert\psi_j\rangle
         \Biggr\|^2_{{\cal H}_{{\rm \scriptscriptstyle PM}}}
        \geq 0 \qquad {{{\rm for}}}\ t\geq 0.
        \end{eqnarray}
Note that according to (\ref{brutta}) the condition in (\ref{cp})
holds provided $t$ is a positive time, thus expressing the
irreversibility of the obtained time evolution, which gives rise to a
semigroup law.

\subsection{The Master Equation}
\label{me}
\noindent
Exploiting the reduction formulas  (\ref{reduction}) and
(\ref{new}), together with (\ref{serve}),
we come to the master equation for the  statistical operator
describing the microsystem
        \begin{eqnarray*}
        {
        d 
        \over
        dt
        } {\varrho}_{kh}
        &=&
        \textrm{Tr}_{{{{\cal H}_{{\rm \scriptscriptstyle PM}}}}}
        \left(
        {\cal L}'
        \left(
        a^{\scriptscriptstyle\dagger}_h a_k 
        \right)
        \rho_{{\rm \scriptscriptstyle PM}}
        \right)
        \\
        &=&
        +
        \frac{i}{\hbar} (E_h-E_k)   \sum_{pq}
        \textrm{Tr}_{{{{\cal H}_{{\rm \scriptscriptstyle PM}}}}}
        \left(
        a^{\scriptscriptstyle\dagger}_h a_k
        a^{\scriptscriptstyle\dagger}_p
        \varrho_{{\rm \scriptscriptstyle M}} a_q \varrho_{pq}
        \right)
        \\
        &&
        -\frac{i}{\hbar} \sum_{pq f}
        \textrm{Tr}_{{{{\cal H}_{{\rm \scriptscriptstyle PM}}}}}
        \left(
        a^{\scriptscriptstyle\dagger}_h
        T{}_f^k(E_k+i\varepsilon) a_f
        a^{\scriptscriptstyle\dagger}_p
        \varrho_{{\rm \scriptscriptstyle M}} a_q \varrho_{pq}
        \right)
        \\
        &&
        +\frac{i}{\hbar} \sum_{pq g}
        \textrm{Tr}_{{{{\cal H}_{{\rm \scriptscriptstyle PM}}}}}
        \left(
        a^{\scriptscriptstyle\dagger}_g
        T{}_g^h{}^{\scriptscriptstyle\dagger}(E_h+i\varepsilon)
        a_k
        a^{\scriptscriptstyle\dagger}_p
        \varrho_{{\rm \scriptscriptstyle M}} a_q \varrho_{pq}
        \right)
        \\
        &&
        +2\frac{\varepsilon}{\hbar}
        \sum_{\lambda\alpha\alpha'\atop pqgf}
        \textrm{Tr}_{{{{\cal H}_{{\rm \scriptscriptstyle PM}}}}}
        \left(
        a^{\scriptscriptstyle\dagger}_g
        \vert\alpha\rangle
        {
        \langle\alpha\vert
        T{}_g^h{}^{\scriptscriptstyle\dagger}(E_h+i\varepsilon)
        \vert\alpha'\rangle
        \over
        E_g + E_{\alpha} -E_h-E_{\alpha'}-i\varepsilon
        }
        \right.
        \\
        &&
        \hphantom{
        +2\frac{\varepsilon}{\hbar}
        \sum_{\lambda\alpha\alpha'\atop pqgf}
        \textrm{Tr}_{{{{\cal H}_{{\rm \scriptscriptstyle PM}}}}}
        \{ =
        }
        \left.
        \times
        {
        \langle\alpha'\vert
        T{}_f^k(E_k+i\varepsilon)
        \vert \lambda \rangle
        \over
        E_f + E_{\lambda} -E_k-E_{\alpha'}+i\varepsilon
        }
        \langle \lambda \vert
        a_f
        a^{\scriptscriptstyle\dagger}_p
        \varrho_{{\rm \scriptscriptstyle M}} a_q \varrho_{pq}
        \right)
        \end{eqnarray*}
which due to     (\ref{zero}) and using the decomposition
$\varrho_{{\rm \scriptscriptstyle M}}=\sum_{\xi}
\pi_{\xi}
\vert {{\xi}} \rangle
{\langle \xi \vert}$ becomes
        \begin{eqnarray*}
        {
        d 
        \over
        dt
        } {\varrho}_{kh}
        &=&
        -\frac{i}{\hbar} (E_k-E_h)
        \varrho_{kh}
        \\
        &&
        -\frac{i}{\hbar} \sum_{f}
        \textrm{Tr}_{{{{\cal H}_{{\rm \scriptscriptstyle PM}}}}}
        \left[
        T{}_f^k(E_k+i\varepsilon)
        \varrho_{{\rm \scriptscriptstyle M}}
        \right]
        \varrho_{fh}
        \\
        &&
        +\frac{i}{\hbar} \sum_{g}
        \varrho_{kg}
        \textrm{Tr}_{{{{\cal H}_{{\rm \scriptscriptstyle PM}}}}}
        \left[
        T{}_g^h{}^{\scriptscriptstyle\dagger}(E_h+i\varepsilon)
        \varrho_{{\rm \scriptscriptstyle M}}
        \right]
        \\
        &&
        +2\frac{\varepsilon}{\hbar}
        \sum_{\eta\xi\atop gf}
        {
        \langle\eta\vert
        T{}_f^k(E_k+i\varepsilon)
        \vert \xi \rangle
        \over
        E_f + E_{\xi} -E_k-E_{\eta}+i\varepsilon
        }
        \pi_\xi
        \varrho_{fg}
        {
        \langle\xi\vert
        T{}_g^h{}^{\scriptscriptstyle\dagger}(E_h+i\varepsilon)
        \vert\eta\rangle
        \over
        E_g + E_{\xi} -E_h-E_{\eta}-i\varepsilon
        }             .
        \end{eqnarray*}

The master equation describing the irreversible time evolution of the  
statistical operator on the chosen time scale can also be written:
        \begin{eqnarray}  
        \label{Lind}  
        {  
        d
        \over  
        dt
        } {\varrho}_{kh}
        &=&  
        -{i \over \hbar}  
        \left(  
        {{E_k}-{E_h}}  
        \right)  
          {\varrho}_{kh}  
          \nonumber \\
        &&
        -  
        {i \over \hbar}  
        \sum_f  {{Q}}_{kf} {\varrho}_{fh}  
        +  
        {i \over \hbar}  
        \sum_g  {\varrho}_{kg} {{Q}}^{*}_{hg}  
        +  
        {1 \over \hbar}  
        \sum_{gf \atop \lambda\xi}
        \left(  
          {{{ L}}_{\lambda\xi}}  
          \right)  
          _{kf}  
        {\varrho}_{fg}  
        {\left( 
        {{{L}}_{\lambda\xi}}  
        \right)}^{*}_{hg} 
        ,  
        \end{eqnarray}  
the quantities appearing in (\ref{Lind}) being defined in the  
following way:  
        \begin{eqnarray}  
        \label{l}  
        {{ Q}}_{kf}  
        &=&  
        \textrm{Tr}_{{{{\cal H}_{{\rm \scriptscriptstyle PM}}}}}  
        \left[  
        {  
        {  
        T{}_{f}^{k}  
        ({E_k}+i{\varepsilon})  
        }  
        {\varrho_{{\rm \scriptscriptstyle M}}}
        }  
        \right]  
        \nonumber  
        \\  
        {{Q}}^{*}_{hg}  
        &=&  
        \textrm{Tr}_{{{{\cal H}_{{\rm \scriptscriptstyle PM}}}}}  
        \left[  
        {  
        T{}_{g}^{h}{}^{\scriptscriptstyle \dagger}  
        (  
        {{E_h}+i{\varepsilon}}  
        )  
        {\varrho_{{\rm \scriptscriptstyle M}}}
        }  
        \right]  
        \nonumber
        \\
        {\left(
          {{{L}}_{\lambda\xi}}  
        \right)}_{kf} 
        &=&  
        \sqrt{2\varepsilon  \pi_\xi}  
        {  
        \langle  
        \lambda  
        \vert  
        {  
        T{}_{f}^{k}  
        ({E_k}+i{\varepsilon})  
        }  
        \vert  
        \xi (t)  
        \rangle  
        \over  
        {{E_f}+{E_{{\xi}}}-{E_k}-E_{\lambda} +i\varepsilon}
        }  
        .  
        \end{eqnarray}  
If we now introduce in ${{{\cal
      H}^1_{{\rm \scriptscriptstyle P}}}}$   the operators  
${\hat {\sf H}_0},  
{\hat {\sf Q}},  
{\hat {{\sf L}}}_{\lambda\xi}$ and ${\hat {\sf \varrho}}$
        \begin{displaymath}
        \langle  
        {g}  
        \vert  
        {\hat {\sf H}}_0  
        \vert  
        {f}  
        \rangle  
        =E_f \delta_{gf}  
        ,  
        \quad  
        \langle  
        {g}  
        \vert  
        {\hat {\sf Q}}  
        \vert  
        {f}  
        \rangle  
        ={Q}_{gf}  
        ,  
        \quad  
        \langle  
        {g}  
        \vert  
        {\hat {{\sf L}}_{\lambda\xi}}  
        \vert  
        {f}  
        \rangle  
        =  
         {\bigl( {{{L}}_{\lambda\xi}} \bigr)}_{gf},  
        \quad  
        \langle  
        {g}  
        \vert  
        {\hat {\sf \varrho}}
        \vert  
        {f}  
        \rangle  
        =  
        \varrho_{gf},  
        \end{displaymath}
eq.\ (\ref{Lind}) takes the operator form:  
        \begin{displaymath}
        {
        d {\hat {\sf \varrho}}
        \over  
                      dt
        }  
        =  
        -{i \over \hbar}  
        \left[{\hat {{\sf H}}}_0  
        +  
        {\hat {\sf V}},
        {\hat {\sf \varrho}}
        \right]
        -{1\over\hbar}  
        \left \{  
        {  
                {\hat {\sf \Gamma}}
                , {\hat {\sf \varrho}}
                }
                \right \}  
        +  
        {1 \over \hbar}  
        \sum^{}_{{\xi\lambda  }}
        {\hat {\sf L}}_{\lambda\xi} {\hat {\varrho}}
        {{\hat {\sf L}}{}_{\lambda\xi}^{\scriptscriptstyle \dagger}}\ ,  
        \end{displaymath}
where
        \begin{displaymath}
        {\hat {\sf V}}=
        {  
        {{\hat {\sf Q}}+  
        {\hat {\sf Q}}{}^{\scriptscriptstyle \dagger}}  
        \over  
        2  
        }  
        , \qquad  
        {\hat {\sf \Gamma}}=i
        {  
        {  
        {\hat {{\sf Q}}}  
        -{\hat {\sf Q}}{}^{\scriptscriptstyle \dagger}}  
        \over  
        2  
        }  
         .  
        \end{displaymath}
Verification of the conservation of the trace of the  
statistical operator leads according to (\ref{N})
to the following relationship  
        \[
        {\hat {\sf \Gamma}}
        =
        {1\over 2}  
        \sum^{}_{{\xi\lambda  }}
        {{\hat {\sf L}}{}_{\lambda\xi}^{\scriptscriptstyle \dagger}}  
        {\hat {\sf L}}{}_{\lambda\xi}  
        ,  
        \]
and therefore to:  
        \begin{equation}  
        \label{n4}  
        {  
        d {\hat {\sf \varrho}}
        \over  
                      dt
        }  
        =  
        -{i \over \hbar}  
        \left[{\hat {{\sf H}}}_0  
        +  
        {\hat {\sf V}},
        {\hat {\sf \varrho}}
        \right]
        -{1\over\hbar}  
        \left \{  
        {  
        {1\over 2}  
        \sum^{}_{{\xi\lambda  }}
        {{\hat {\sf L}}{}_{\lambda\xi}^{\scriptscriptstyle \dagger}}  
        {\hat {\sf L}}{}_{\lambda\xi}  
                , {\hat {\sf \varrho}}
                }
                \right \}  
        +  
        {1 \over \hbar}  
        \sum^{}_{{\xi\lambda  }}
        {\hat {\sf L}}_{\lambda\xi} {\hat {\varrho}}
        {{\hat {\sf L}}{}_{\lambda\xi}^{\scriptscriptstyle  
        \dagger}}\ .  
        \end{equation}  
We have thus obtained a general structure of master-equation for the
subdynamics of the microsystem, which we shall now apply to some
specific example, making a suitable Ansatz for the T-matrix appearing
in (\ref{l}) and describing the collisions which drive the interaction
between microsystem and macrosystem. This matrix, obtained averaging
over the state of the macrosystem the matrix
(\ref{a13}), operator-valued in ${{{\cal
      H}_{{\rm \scriptscriptstyle M}}}}$, keeps the statistical
mechanics properties of the macrosystem into account and is a natural
place for fundamental or phenomenological Ansatz.

\section{Master Equation for a Test Particle in a Quantum Gas}
\label{test}
\noindent
We now aim to apply the master-equation (\ref{n4}) to the case of a
test particle interacting through collisions with a quantum fluid
considered in Ref.\cite{art4}, a physical example corresponding to
the so-called Rayleigh gas.\cite{Spohn} Exploiting the fact that
interactions at microphysical level are translation invariant, a
general Ansatz for the T-matrix (\ref{212a}) is given by
\begin{equation}
        T{}_{h}^{k} \left( z \right) 
        =
        {\int_\omega d^3 \! {{\bf x}} \,}
        {\int_\omega d^3 \! {{\bf y}} \,}  
        \psi^{\scriptscriptstyle\dagger}({{\bf x}})  
        u_k^{*}({{\bf y}})  
        t(z,{{\bf x}}-{{\bf y}})  
        u_h({{\bf y}})  
        \psi({{\bf x}})  ,
\label{100}
\end{equation}
where the integrals are extended over the region in which the system
is confined. We now only want to consider local dissipation effects, so 
that we will suppose the system to be homogeneous and use as quantum numbers
momentum eigenvalues. This picture holds provided we are
sufficiently far away from the boundaries and the peculiar features of 
the normal modes do not play a relevant role, which for a real
confined system will generally be true only for a finite
time. According to this picture at a later stage we will take a
thermodynamic or continuum limit, thus obtaining an expression
describing an idealized situation in which the confinement is
completely removed, actual calculations are made easier through the
introduction of integrals instead of sums and invariance properties
with respect to symmetry transformations are more directly formulated
and checked.\cite{art5}

Using as quantum numbers momentum eigenvalues and introducing creation
and destruction operators $b^{\scriptscriptstyle\dagger}_\eta$,
$b_\mu$ in the Fock-space of the macrosystem ${{{\cal H}_{{\rm
        \scriptscriptstyle M}}}}$ one obtains from (\ref{100}) the
following expression in terms of the Fourier transform of the
translation and rotation invariant interaction kernel, which therefore
only depends on the modulus of the momentum transfer
        \begin{equation}
        \label{101}
        T{}_{h}^{k} 
        =  
        \sum_{\eta\mu}
        \delta_{p_\eta +p_k,p_h+p_\mu}
        \tilde{t} (
        |
        {{\bf p}}_{\mu}-{{\bf p}}_{\eta}
        |
        )
        b^{\scriptscriptstyle\dagger}_{\eta}
        b_\mu                      ,
        \end{equation}
and where the slow energy dependence of the T-matrix has been
neglected for simplicity, this in turn implying that the potential
term in (\ref{n4}), expressible in terms of the forward scattering
amplitude,\cite{art3} is in fact a c-number, of relevance only for
wave-like, coherent dynamics.\cite{art2} Eq.~(\ref{n4}) then becomes
\begin{equation}
  \label{102}
        {  
        d {\hat \varrho}  
        \over  
                      dt
        }  
        =
        -
        {i \over \hbar}
        [{\hat {{\sf H}}}_0
        ,
        {\hat \varrho}
        ]
        +
        {\cal L} [{\hat \varrho}],
\end{equation}
where ${\hat {{\sf H}}}_0$ is the Hamiltonian for the free particle
${\hat {{\sf H}}}_0= {\hat {\sf p}^2 \over 2M }$ and according to (\ref{101})
\begin{eqnarray}
\label{pra1}
  {\cal L} [{\hat \varrho}]
&=&
{}+\frac{2\varepsilon}{\hbar}
        \sum^{}_{{\lambda\xi}}
        \sum^{}_{kf}
        \sum^{}_{hg}
\vert {{\bf p}}_f \rangle
\frac
{
        \sum_{\eta\mu}
        \delta_{p_\eta +p_f,p_k+p_\mu}
        \tilde{t} (
        |
        {{\bf p}}_{\mu}-{{\bf p}}_{\eta}
        |
        )
        \langle
        \lambda
        \vert
        b^{\scriptscriptstyle\dagger}_{\eta}
        b_\mu  
        \vert
        \xi
        \rangle
}
{
{{E_k}-{E_f}+{E}_{\xi}-{E_{{\lambda}}} +i\varepsilon}
}
\nonumber
\\
&&
\hphantom{\frac{2\varepsilon}{\hbar}
        }
\times
        \langle {{\bf p}}_k \vert
        {\hat \varrho}
        \vert {{\bf p}}_h \rangle
\pi_\xi
\frac
{
        \sum_{\eta'\mu'}
        \delta_{p_{\eta'} +p_g,p_h+p_{\mu'}}
        \tilde{t}^* (
        |
        {{\bf p}}_{\mu'}-{{\bf p}}_{\eta'}
        |
        )
        \langle
        \xi
        \vert
        b^{\scriptscriptstyle\dagger}_{\mu'}
        b_{\eta'}  
        \vert
        \lambda
        \rangle
}
{
{{E_h}-{E_g}+{E_{{\xi}}}-{E}_{\lambda} -i\varepsilon}
}
\langle {{\bf p}}_g \vert
\nonumber
\\
&&
{}-\frac{\varepsilon}{\hbar}
        \sum^{}_{{\lambda\xi}}
        \sum^{}_{k}
        \sum^{}_{fg}
\{
\vert {{\bf p}}_f \rangle         
\langle {{\bf p}}_g \vert
,
{\hat \varrho}
\}
\frac
{
        \sum_{\eta\mu}
        \delta_{p_\eta +p_k,p_g+p_\mu}
        \tilde{t} (
        |
        {{\bf p}}_{\mu}-{{\bf p}}_{\eta}
        |
        )
        \langle
        \lambda
        \vert
        b^{\scriptscriptstyle\dagger}_{\eta}
        b_\mu  
        \vert
        \xi
        \rangle
}
{
{{E_f}-{E_k}+{E}_{\xi}-{E}_{\lambda} -i\varepsilon}
}
\nonumber
\\
&&
\hphantom{\frac{2\varepsilon}{\hbar}
}
\times
\pi_\xi
\frac
{
        \sum_{\eta'\mu'}
        \delta_{p_{\eta'} +p_k,p_f+p_{\mu'}}
        \tilde{t}^* (
        |
        {{\bf p}}_{\mu'}-{{\bf p}}_{\eta'}
        |
        )
        \langle
        \xi
        \vert
        b^{\scriptscriptstyle\dagger}_{\mu'}
        b_{\eta'}  
        \vert
        \lambda
        \rangle
}
{
{{E_g}-{E_k}+{E}_{\xi}-{E}_{\lambda} +i\varepsilon}
}.
\end{eqnarray}
Due to translation invariance of the interaction it is now natural and 
convenient to introduce as variables the momentum  transfers 
$
{{{\bf q}}}={{\bf p}}_{\mu}-{{\bf p}}_{\eta}
$,
$
{{{\bf q}}}'={{\bf p}}_{\mu'}-{{\bf p}}_{\eta'}
$
and accordingly the operators $\rho_q$
\begin{equation}
\label{rq}
  \rho_q =         
        \sum_{\mu}
        b^{\scriptscriptstyle\dagger}_{\mu}
        b_{\mu+q},
\end{equation}
so that (\ref{pra1}) becomes 
\begin{eqnarray*}
  {\cal L} [{\hat \varrho}]
&=&
{}+\frac{2\varepsilon}{\hbar}
        \sum^{}_{{\lambda\xi}}
        \sum^{}_{pp'}
        \sum^{}_{qq'}{}'
        \tilde{t} (q) \tilde{t}^* (q') 
        e^{{i\over\hbar}{{\bf q}}\cdot{\hat {{\sf x}}}}
        \vert {{\bf p}} \rangle
        \langle {{\bf p}} \vert
        {\hat \varrho}
        \vert {{\bf p}}' \rangle
        \langle {{\bf p}}' \vert
        e^{-{i\over\hbar}{{\bf q}}'\cdot{\hat {{\sf x}}}}
\\
&&
\nonumber
\hphantom{\frac{2\varepsilon}{\hbar}
        \sum^{}_{{\lambda\xi}}
}
\times
\frac
{
1
}
{
{{E_p}-{E_{p+q}}+{E}_{\xi}-{E_{{\lambda}}} +i\varepsilon}
}
\frac
{
1
}
{
{{E_{p'}}-{E_{p'+q'}}+{E}_{\xi}-{E_{{\lambda}}}-i\varepsilon}
}
\\
&&
\nonumber
\hphantom{pippo}
\\
&&
\nonumber
\hphantom{\frac{2\varepsilon}{\hbar}
        \sum^{}_{{\lambda\xi}}
ancora}
\times
\langle\lambda\vert\rho_q\vert\xi\rangle
\pi_\xi 
\langle\xi\vert\rho_{q'}^{\scriptscriptstyle\dagger}\vert\lambda\rangle
\\
&&
\nonumber
{}-\frac{\varepsilon}{\hbar}
        \sum^{}_{{\lambda\xi}}
        \sum^{}_{p}
        \sum^{}_{qq'}{}'
        \tilde{t} (q) \tilde{t}^* (q') 
\{
\vert {{\bf p}} \rangle         
\langle {{\bf p}}+ {{\bf q}}'- {{\bf q}} \vert
,
{\hat \varrho}
\}
\\
&&
\nonumber
\hphantom{\frac{2\varepsilon}{\hbar}
        \sum^{}_{{\lambda\xi}}
}
\times
\frac
{
1
}
{
{{E_p}-{E_{p+q'}}+{E}_{\xi}-{E_{{\lambda}}} -i\varepsilon}
}
\frac
{
1
}
{
{{E_{p+q'-q}}-{E_{p+q'}}+{E}_{\xi}-{E_{{\lambda}}}+i\varepsilon}
}
\\
&&
\nonumber
\hphantom{pippo}
\\
&&
\nonumber
\hphantom{\frac{2\varepsilon}{\hbar}
        \sum^{}_{{\lambda\xi}}
ancora}
\times
\langle\lambda\vert\rho_q\vert\xi\rangle
\pi_\xi 
\langle\xi\vert\rho_{q'}^{\scriptscriptstyle\dagger}\vert\lambda\rangle
,
\end{eqnarray*}
where the contributions for ${{\bf q}}={{\bf q}}'=0$ have canceled
out, as denoted by the primed sum. Expressing the denominators through
a Laplace transform and denoting the energy transfer $E_{p+q} - E_p$
by $\Delta E_q ({{\bf p}})$, the ensemble average over
${{\varrho}_{{\rm \scriptscriptstyle M}}}$ by $ \left \langle \ldots
\right \rangle $, the Heisenberg operator $ e^{+{i\over\hbar} H_{{\rm
      \scriptscriptstyle M}}t} \rho_q e^{-{i\over\hbar} H_{{\rm
      \scriptscriptstyle M}}t} $ by $\rho_q (t)$, we have
\begin{eqnarray*}
  {\cal L} [{\hat \varrho}]
&=&
{}+\frac{2\varepsilon}{\hbar}
         \sum^{}_{pp'}
        \sum^{}_{qq'}{}'
        \tilde{t} (q) \tilde{t}^* (q') 
        e^{{i\over\hbar}{{\bf q}}\cdot{\hat {{\sf x}}}}
        \vert {{\bf p}} \rangle
        \langle {{\bf p}} \vert
        {\hat \varrho}
        \vert {{\bf p}}' \rangle
        \langle {{\bf p}}' \vert
        e^{-{i\over\hbar}{{\bf q}}'\cdot{\hat {{\sf x}}}}
\\
&&
\hphantom{\frac{2\varepsilon}{\hbar}
}
\times
        {1\over\hbar^2}
         \int_0^\infty  
        \!  
        d\tau \, e^{-{\varepsilon\over\hbar}\tau}  
         \!\int_0^\infty  
        \!  
        d\tau' \, e^{-{\varepsilon\over\hbar}\tau'}  
        \,  
e^{-{i\over\hbar}\Delta E_q ({{\bf p}}) \tau}
e^{+{i\over\hbar}\Delta E_{q'} ({{\bf p}}') \tau'}
\\
&&
\hphantom{\frac{2\varepsilon}{\hbar}
        {1\over\hbar^2}
         \int_0^\infty  
        \!  
        d\tau \, e^{-{\varepsilon\over\hbar}\tau}  
         \!\int_0^\infty  
        \!  
        d\tau' \,
}
\times
\langle
\rho_{q'}^{\scriptscriptstyle\dagger}\rho_q (\tau-\tau')
\rangle
\nonumber
\\
&&
{}-\frac{\varepsilon}{\hbar}
         \sum^{}_{p}
        \sum^{}_{qq'}{}'
        \tilde{t} (q) \tilde{t}^* (q') 
\{
\vert {{\bf p}} \rangle         
\langle {{\bf p}}+ {{\bf q}}'- {{\bf q}} \vert
,
{\hat \varrho}
\}
\\
&&
\hphantom{\frac{2\varepsilon}{\hbar}
}
\times
        {1\over\hbar^2}
         \int_0^\infty  
        \!  
        d\tau \, e^{-{\varepsilon\over\hbar}\tau}  
         \!\int_0^\infty  
        \!  
        d\tau' \, e^{-{\varepsilon\over\hbar}\tau'}  
        \,  
e^{+{i\over\hbar}\Delta E_{q'} ({{\bf p}}) \tau}
e^{-{i\over\hbar}\Delta E_{q} ({{\bf p}}+ {{\bf q}}'- {{\bf q}}) \tau'}
\\
&&
\hphantom{\frac{2\varepsilon}{\hbar}
        {1\over\hbar^2}
         \int_0^\infty  
        d\tau \, e^{-{\varepsilon\over\hbar}\tau}  
         \!\int_0^\infty  
        \!  
        d\tau' \,
}
\times
\langle
\rho_{q'}^{\scriptscriptstyle\dagger}\rho_q (\tau-\tau')
\rangle
.  
\end{eqnarray*}
Using the identity $1= \int dt \, \delta (t-\alpha)$ and giving for
the Dirac $\delta$ a Fourier representation one obtains
\begin{eqnarray*}
  {\cal L} [{\hat \varrho}]
&=&
{}+\frac{2\varepsilon}{\hbar}
         \sum^{}_{pp'}
        \sum^{}_{q}{}'
        |\tilde{t} (q)|^2
        e^{{i\over\hbar}{{\bf q}}\cdot{\hat {{\sf x}}}}
        \vert {{\bf p}} \rangle
        \langle {{\bf p}} \vert
        {\hat \varrho}
        \vert {{\bf p}}' \rangle
        \langle {{\bf p}}' \vert
        e^{-{i\over\hbar}{{\bf q}}\cdot{\hat {{\sf x}}}}
\\
&&
\hphantom{\frac{2\varepsilon}{\hbar}
}
\times
        {1\over\hbar^2}
         \int_0^\infty  
        \!  
        d\tau \, e^{-{\varepsilon\over\hbar}\tau}  
         \!\int_0^\infty  
        \!  
        d\tau' \, e^{-{\varepsilon\over\hbar}\tau'}  
\!\int dE         \,
e^{-{i\over\hbar}[\Delta E_q ({{\bf p}}) - E] \tau}
e^{+{i\over\hbar}[\Delta E_{q} ({{\bf p}}') -E] \tau'}
\\
&&
\hphantom{
\frac{2\varepsilon}{\hbar}
\times
}
\times        {  
        1  
        \over  
         2\pi\hbar
        }  
        \int dt \, 
        e^{
        {
        i
        \over
         \hbar
        }
        E t
        }  
\langle
\rho_{q}^{\scriptscriptstyle\dagger}\rho_q (t)
\rangle
\nonumber
\\
&&
{}-\frac{\varepsilon}{\hbar}
         \sum^{}_{p}
        \sum^{}_{q}{}'
        |\tilde{t} (q)|^2
\{
\vert {{\bf p}} \rangle         
\langle {{\bf p}} \vert
,
{\hat \varrho}
\}
\\
&&
\hphantom{\frac{2\varepsilon}{\hbar}
}
\times
        {1\over\hbar^2}
         \int_0^\infty  
        \!  
        d\tau \, e^{-{\varepsilon\over\hbar}\tau}  
         \!\int_0^\infty  
        \!  
        d\tau' \, e^{-{\varepsilon\over\hbar}\tau'}  
\!\int dE         \,
e^{-{i\over\hbar}[\Delta E_q ({{\bf p}}) -E] \tau}
e^{+{i\over\hbar}[\Delta E_{q} ({{\bf p}})-E] \tau'}
\\
&&
\hphantom{\frac{2\varepsilon}{\hbar}
\times}
\times        {  
        1  
        \over  
         2\pi\hbar
        }  
        \int dt \, 
        e^{
        {
        i
        \over
         \hbar
        }
        E t
        }  
\langle
\rho_{q}^{\scriptscriptstyle\dagger}\rho_q (t)
\rangle
,  
\end{eqnarray*}
where a major simplification has been given by the homogeneity of the
macrosystem, implying ${{{\bf q}}}={{{\bf q}}}'$. We can now undo
the Laplace transform coming to
\begin{eqnarray*}
  {\cal L} [{\hat \varrho}]
&=&
{}+\frac{2\varepsilon}{\hbar}
         \sum^{}_{pp'}
        \sum^{}_{q}{}'
        |\tilde{t} (q)|^2
        e^{{i\over\hbar}{{\bf q}}\cdot{\hat {{\sf x}}}}
        \vert {{\bf p}} \rangle
        \langle {{\bf p}} \vert
        {\hat \varrho}
        \vert {{\bf p}}' \rangle
        \langle {{\bf p}}' \vert
        e^{-{i\over\hbar}{{\bf q}}\cdot{\hat {{\sf x}}}}
\\
&&
\hphantom{\frac{2\varepsilon}{\hbar}
        \sum^{}_{kf}
        \sum^{}_{hg}}
\times
\int dE         \,
\frac{\varepsilon}{\pi}
\frac{1}{E-\Delta E_q ({{\bf p}}) +i\varepsilon}
\frac{1}{E-\Delta E_q ({{\bf p}}') -i\varepsilon}
\\
&&
\hphantom{\frac{2\varepsilon}{\hbar}
        \sum^{}_{kf}
        \sum^{}_{hg}\times}
\times
        {  
        1  
        \over  
         2\pi\hbar
        }  
        \int dt \, 
        e^{
        {
        i
        \over
         \hbar
        }
        E t
        }  
\langle
\rho_{q}^{\scriptscriptstyle\dagger}\rho_q (t)
\rangle
\nonumber
\\
&&
{}-\frac{\varepsilon}{\hbar}
         \sum^{}_{p}
        \sum^{}_{q}{}'
        |\tilde{t} (q)|^2
\{
\vert {{\bf p}} \rangle         
\langle {{\bf p}} \vert
,
{\hat \varrho}
\}
\\
&&
\hphantom{\frac{2\varepsilon}{\hbar}
        \sum^{}_{kf}
        \sum^{}_{hg}}
\times
\int dE         \,
\frac{\varepsilon}{\pi}
\frac{1}{E-\Delta E_q ({{\bf p}}) -i\varepsilon}
\frac{1}{E-\Delta E_q ({{\bf p}}) +i\varepsilon}
 \\
&&
\hphantom{\frac{2\varepsilon}{\hbar}
        \sum^{}_{kf}
        \sum^{}_{hg}\times}
\times
       {  
        1  
        \over  
         2\pi\hbar
        }  
        \int dt \, 
        e^{
        {
        i
        \over
         \hbar
        }
        E t
        }  
\langle
\rho_{q}^{\scriptscriptstyle\dagger}\rho_q (t)
\rangle
.  
\end{eqnarray*}
As a last step we exploit the quasi-diagonality of ${\hat \varrho}$ in 
this representation , linked to its slow variability, and substitute
in the denominators of the first term ${{\bf p}}$, ${{\bf p}}'$
with the symmetric expression $\frac 12 ({{\bf p}} +
{{\bf p}}')$, so that we obtain the expression 
\begin{eqnarray}
  \label{pra3}
  {\cal L} [{\hat \varrho}]
&=& \nonumber
{}+\frac{2\varepsilon}{\hbar}
         \sum^{}_{pp'}
        \sum^{}_{q}{}'
        |\tilde{t} (q)|^2
        e^{{i\over\hbar}{{\bf q}}\cdot{\hat {{\sf x}}}}
        \vert {{\bf p}} \rangle
        \langle {{\bf p}} \vert
        {\hat \varrho}
        \vert {{\bf p}}' \rangle
        \langle {{\bf p}}' \vert
        e^{-{i\over\hbar}{{\bf q}}\cdot{\hat {{\sf x}}}}
\\
&&\nonumber
\hphantom{\frac{2\varepsilon}{\hbar}
        \sum^{}_{kf}
        \sum^{}_{hg}}
\times
\int dE         \,
\delta \left( E-\Delta E_q \left(\frac {{{\bf p}} + {{\bf
          p}}'}{2}\right) \right) 
\\
&&\nonumber
\hphantom{\frac{2\varepsilon}{\hbar}
        \sum^{}_{kf}
        \sum^{}_{hg}\times}
\times
        {  
        1  
        \over  
         2\pi\hbar
        }  
        \int dt \, 
        e^{
        {
        i
        \over
         \hbar
        }
        E t
        }  
\langle
\rho_{q}^{\scriptscriptstyle\dagger}\rho_q (t)
\rangle
\nonumber
\\
&&\nonumber
{}-\frac{\varepsilon}{\hbar}
         \sum^{}_{p}
        \sum^{}_{q}{}'
        |\tilde{t} (q)|^2
\{
\vert {{\bf p}} \rangle         
\langle {{\bf p}} \vert
,
{\hat \varrho}
\}
\\
&&\nonumber
\hphantom{\frac{2\varepsilon}{\hbar}
        \sum^{}_{kf}
        \sum^{}_{hg}}
\times
\int dE         \,
\delta \left( E-\Delta E_q ({{\bf p}}) \right)
 \\
&&
\hphantom{\frac{2\varepsilon}{\hbar}
        \sum^{}_{kf}
        \sum^{}_{hg}\times}
\times
       {  
        1  
        \over  
         2\pi\hbar
        }  
        \int dt \, 
        e^{
        {
        i
        \over
         \hbar
        }
        E t
        }  
\langle
\rho_{q}^{\scriptscriptstyle\dagger}\rho_q (t)
\rangle
.  
\end{eqnarray}
Apart form a factor $N$ corresponding to the total number of particles 
in the macroscopic system the two-point correlation function appearing in
(\ref{pra3}) is the well-known
dynamic structure factor,\cite{Lovesey,Griffin} here given in terms of 
momentum ${{\bf q}}$ and energy $E$ transferred to the particle in
the collision
\begin{equation}
  \label{qq}
        S ({{\bf q}},E)=
        {  
        1  
        \over  
         2\pi\hbar
        }  
        {
        1
        \over
         N
        }
        \int dt \, 
        e^{
        {
        i
        \over
         \hbar
        }
        E t 
        }  
         \langle  
          \rho_q^{\scriptscriptstyle\dagger} \rho_q (t)
         \rangle
         ,
\end{equation}
and evaluated in (\ref{pra3}) for energy transfers $\Delta E_q (\frac
{{{\bf p}} + 
{{\bf p}}'}{2} )$ and $\Delta E_q ({{\bf p}})$. The master-equation
(\ref{102}) obtained from (\ref{n4}) through the Ansatz (\ref{101})
can therefore be generally expressed in terms of the
dynamic structure factor of the macrosystem through:
\begin{eqnarray}
        \label{pippo}
        {  
        d {\hat \varrho}  
        \over  
                      dt
        }  
        &=&
        -
        {i \over \hbar}
        [
        {\hat {{\sf H}}}_0
        ,
        {\hat \varrho}
        ]
        +
        {\cal L} [{\hat \varrho}]
        \nonumber
        \\
        &=&
        -
        {i \over \hbar}
        [
        {\hat {{\sf H}}}_0
        ,
        {\hat \varrho}
        ]
        \nonumber
        \\
        \nonumber
        &&+
        {2\pi \over\hbar}N
        \sum_{q}{}'
        {
        | \tilde{t} (q) |^2
        }
        \Biggl[
        \sum_{p p'}
        e^{{i\over\hbar}{{\bf q}}\cdot{\hat {{\sf x}}}}
        \vert {{\bf p}} \rangle
        \langle {{\bf p}} \vert
        {\hat \varrho}
        \vert {{\bf p}}' \rangle
        \langle {{\bf p}}' \vert
        e^{-{i\over\hbar}{{\bf q}}\cdot{\hat {{\sf x}}}}
        S\left( {{\bf q}},\Delta E_q \left(\frac {{{\bf p}} + {{\bf
                  p}}'}{2} \right) \right) 
        \nonumber
        \\
        &&
        \hphantom{
        {  
        d {\hat \varrho}  
        \over  
                      dt
        }  
        =
        +
        {2\pi \over\hbar}
        \sum_{q}{}'
        {
        | \tilde{t} (q) |^2
        }
        }
        - {1\over 2}
        \sum_{p}
        \left \{
        \vert {{\bf p}} \rangle
        \langle {{\bf p}} \vert,
        {\hat \varrho}
        \right \}
        S\left( {{\bf q}},\Delta E_q ({{\bf p}}  ) \right)
        \Biggr]
        .
        \end{eqnarray}
In particular, provided an approximation of the form
\begin{equation}
        \label{x}
        S_{\rm \scriptscriptstyle}\left({{\bf q}},
        {
        E+E' \over 2
        }\right)
        \approx
        \sqrt{
        S_{\rm \scriptscriptstyle}({{\bf q}},E)
        }
        \sqrt{
        S_{\rm \scriptscriptstyle}({{\bf q}},E')
        }
        \end{equation}
can be assumed, eq.~(\ref{pippo}) retains a Lindblad structure,
typical of the generators of completely positive time
evolutions,\cite{Lindblad,HolevoNEW} which in the continuum limit is
given by
\begin{eqnarray}
  \label{pregeneral}
        {  
        d {\hat \varrho}  
        \over  
                      dt
        }  
        &=&
        -
        {i \over \hbar}
        [
        {\hat {{\sf H}}}_0
        ,
        {\hat \varrho}
        ]
        +
        {\cal L} [{\hat \varrho}]
        \nonumber
        \\
        &=&
        -
        {i \over \hbar}
        \left[
        {
        {\hat {{\sf p}}}^2
        \over
        2M
        }
        ,
        {\hat \varrho}
        \right]
        +
        {2\pi \over\hbar}
        (2\pi\hbar)^3
        n
        \int d^3\!
        {{\bf q}}
        \,  
        {
        | \tilde{t} (q) |^2
        }
      \nonumber
      \\
        &&
        \hphantom{}
        \times
      \Biggl[
        e^{{i\over\hbar}{{\bf q}}\cdot{\hat {{\sf x}}}}
        \sqrt{
        S({{\bf q}},\Delta E_q ({\hat {{\sf p}}}))
        }
        {\hat \varrho}
        \sqrt{
        S({{\bf q}},\Delta E_q ({\hat {{\sf p}}}))
        }
        e^{-{i\over\hbar}{{\bf q}}\cdot{\hat {{\sf x}}}}
        -
        \frac 12
        \left \{
        S({{\bf q}},\Delta E_q ({\hat {{\sf p}}})),
        {\hat \varrho}
        \right \}
        \Biggr].
        \nonumber
        \\
\end{eqnarray}
Exploiting the fact that
\begin{displaymath}
  \Delta E_q ({\bf p})=E_{p+q} - E_p=        {
        q^2
        \over
             2M
        }
        +
        {{{\bf q}}\cdot{{\bf p}} \over M}
\end{displaymath}
we will use in the following the equivalent notations
\begin{equation}
  \label{ss}
  S({\bf q},E)\equiv S({\bf q},\Delta E_q ({\bf p}))\equiv S({\bf q},{\bf p}),
\end{equation}
so that one can put in major evidence in the master-equation the
relevant quantities: momentum transfer ${\bf q}$ and the operators
position and momentum for the microsystem ${\hat {{\sf x}}}$ and
${\hat {{\sf p}}}$. In particular the dynamic structure factor is
operator-valued due to its dependence on the momentum operator for the
microsystem ${\hat {{\sf p}}}$. According to (\ref{ss})
eq.~(\ref{pregeneral}) therefore becomes
\begin{eqnarray}
  \label{general}
        {  
        d {\hat \varrho}  
        \over  
                      dt
        }  
        &=&
        -
        {i \over \hbar}
        [
        {\hat {{\sf H}}}_0
        ,
        {\hat \varrho}
        ]
        +
        {\cal L} [{\hat \varrho}]
        \nonumber
        \\
        &=&
        -
        {i \over \hbar}
        \left[
        {
        {\hat {{\sf p}}}^2
        \over
        2M
        }
        ,
        {\hat \varrho}
        \right]
        +
        {2\pi \over\hbar}
        (2\pi\hbar)^3
        n
        \int d^3\!
        {{\bf q}}
        \,  
        {
        | \tilde{t} (q) |^2
        }
      \nonumber
      \\
        &&
        \hphantom{cosicosi}
        \times
      \Biggl[
        e^{{i\over\hbar}{{\bf q}}\cdot{\hat {{\sf x}}}}
        \sqrt{
        S({{\bf q}},{\hat {{\sf p}}})
        }
        {\hat \varrho}
        \sqrt{
        S({{\bf q}},{\hat {{\sf p}}})
        }
        e^{-{i\over\hbar}{{\bf q}}\cdot{\hat {{\sf x}}}}
        -
        \frac 12
        \left \{
        S({{\bf q}},{\hat {{\sf p}}}),
        {\hat \varrho}
        \right \}
        \Biggr].
\end{eqnarray}

Eq.~(\ref{general}) is one of the main results presented in this
paper, giving a general structure of master-equation driving a
completely positive time evolution for a test particle interacting
through collisions with a fluid. In its expression only quantities
with a direct physical meaning appear: the square modulus of the
Fourier transform of the T-matrix, stating the relevance of the single
collisions, and the dynamic structure factor, which accounts for the
statistical mechanics properties of the macrosystem, giving its
response to external perturbations; both together, according to
(\ref{diff}), essentially give the scattering rate, as appropriate in a
master-equation. The two-point correlation function $S$ defined in
(\ref{qq}) appears 
operator-valued, as to be expected in a quantum framework, thus
determining the particular structure (\ref{general}). In fact
(\ref{general}),  as well as the general Lindblad
structure,\cite{Lindblad} would become meaningless if all operators
appearing in it were c-numbers, thus all commuting and therefore
loosing their distinctive quantum feature.

We note in passing that (\ref{general}) gives a physical example (to
our knowledge the first one) of a
recent general mathematical result on the structure of generators of
translation covariant quantum dynamical semigroups.\cite{HolevoTI}
The validity of the approximation (\ref{x}), which is in any case
well-defined because the dynamic structure factor is always a positive 
function, being related to the scattering cross-section as
shown in (\ref{diff}), depends both on the energy dependence of
the dynamic structure factor and on the quasi-diagonality of the
statistical operator describing the microsystem, which can be safely
assumed if the microsystem is not too far from equilibrium and
correspondingly its dynamics is on a not too short time scale. For a
discussion of this point in a specific physical application see
Sec.~4\ref{cpqbm} and also
Ref.\cite{Axel}, where a master-equation of the form (\ref{pippo})
was considered for a free Bose or Maxwell-Boltzmann gas, as can be
recognized keeping (\ref{abf}) and  (\ref{amb}) into account. In any
case the neglected terms are at least quadratic in the energy difference.

\subsection{Dynamic Structure Factor}
\label{structure-factor}
\noindent
Let us now go back to the physical meaning of the
dynamic structure factor. This two-point correlation function
translates into the master-equation driving the subdynamics of the
microsystem the statistical mechanics properties of the macrosystem,
giving in particular its spectrum of spontaneous fluctuations. The
relevance of the dynamic structure factor mainly lies in its direct
experimental access, in fact as first shown by van Hove\cite{vanHove}
it appears in the expression of the energy dependent scattering cross-section 
of a microscopic probe off the considered system, which gives the
physical reason for its being always positive: in particular one has
the result
\begin{equation}
  \label{diff}
  \frac{d^2 \sigma}{d\Omega_{p'} dE_{p'}}=
\left({2\pi\hbar}\right)^6
\left(\frac{M}{2\pi\hbar^2}\right)^2
\frac{p'}{p}
        {
        | \tilde{t} (q) |^2
        }
  S ({{\bf q}},E)
,
\end{equation}
which gives the scattering cross-section per target particle, if the
momentum of the incoming probe changes from ${{\bf p}}$ to
${{\bf p}}' = {{\bf p}}+{{\bf q}}$. 

The dynamic structure factor, apart from being accessible through
experiments and a natural starting point for phenomenological
expressions, can be exactly calculated in the case of a collection of
free particles. In fact from the general expression (\ref{qq}) one
has, denoting with $b^{\scriptscriptstyle\dagger}_\eta$, $b_\mu$
creation and destruction operators associated to the modes of the
macrosystem
\begin{eqnarray*}
        S ({{\bf q}},E)&=&
        {  
        1  
        \over  
         2\pi\hbar
        }  
        {
        1
        \over
         N
        }
        \int dt \, 
        e^{
        {
        i
        \over
         \hbar
        }
        E t 
        }  
         \langle  
          \rho_q^{\scriptscriptstyle\dagger} \rho_q (t)
         \rangle
         \\
         &=&
        {  
        1  
        \over  
         2\pi\hbar
        }  
        {
        1
        \over
         N
        }
        \int dt \, 
        e^{
        {
        i
        \over
         \hbar
        }
        E t 
        }  
        \sum_{\mu,\eta}
         \langle  
         b^{\scriptscriptstyle\dagger}_{\mu}
        b_{\mu-q}b^{\scriptscriptstyle\dagger}_{\eta} (t)
        b_{\eta+q} (t)
         \rangle
         \\
         &=&
        {  
        1  
        \over  
         2\pi\hbar
        }  
        {
        1
        \over
         N
        }
        \int dt \, 
        e^{
        {
        i
        \over
         \hbar
        }
        E t 
        }  
        \sum_{\mu,\eta}\textrm{Tr}_{{\cal H}_{\rm \scriptscriptstyle M}}
         \left(
         \frac{e^{-\beta H_{\rm \scriptscriptstyle M}}}{{\cal Z}}  
         b^{\scriptscriptstyle\dagger}_{\mu}
        b_{\mu-q}
        e^{+\frac{i}{\hbar} H_{\rm \scriptscriptstyle M}t}
        b^{\scriptscriptstyle\dagger}_{\eta}b_{\eta+q} 
        e^{-\frac{i}{\hbar} H_{\rm \scriptscriptstyle M}t}
        \right)
         ,
\end{eqnarray*}
and for free particles, setting $\mu\rightarrow p_{\mu}$, $\eta\rightarrow
p_{\eta}$, $H_{\rm \scriptscriptstyle M}\rightarrow \sum_\mu{{{\bf
      p}}_{\mu}^2\over 2m} 
b^{\scriptscriptstyle\dagger}_{\mu} b_{\mu}$, where $m$ is the mass of the
particles, one has
\begin{displaymath}
        S ({{\bf q}},E)=
        {  
        1  
        \over  
         2\pi\hbar
        }  
        {
        1
        \over
         N
        }
         \sum_{\mu}
         \int dt \, 
        e^{
        {
        i
        \over
         \hbar
        }
        \left( E+ {{{\bf p}}_{\mu}^2\over 2m}-{({{{\bf p}}_{\mu} -
              {{\bf q}}})^2\over 2m}\right)t  
        }  
         \sum_{\eta} \langle  
         b^{\scriptscriptstyle\dagger}_{\mu}
        b_{\mu-q}b^{\scriptscriptstyle\dagger}_{\eta}
        b_{\eta+q} 
         \rangle_0 
         ,
\end{displaymath}
where $ \langle \ldots  \rangle_0$ denotes the expectation value
calculated with the free Hamiltonian. Using Wick's theorem at finite
temperature one then has
\begin{eqnarray*}
\nonumber
         \sum_{\eta} \langle  
         b^{\scriptscriptstyle\dagger}_{\mu}
        b_{\mu-q}b^{\scriptscriptstyle\dagger}_{\eta}
        b_{\eta+q} 
         \rangle_0 
         &=&
        \langle n_\mu  \rangle_0  \pm 
         \sum_{\eta} \langle  
         b^{\scriptscriptstyle\dagger}_{\mu}
        b^{\scriptscriptstyle\dagger}_{\eta}
        b_{\mu-q} b_{\eta+q}
         \rangle_0 
         \\
         \nonumber
         &=&
        \langle n_\mu  \rangle_0  \pm 
         \sum_{\eta} 
         \{
         \langle  
        b^{\scriptscriptstyle\dagger}_{\eta}
        b_{\mu-q}
         \rangle_0 
         \langle  b^{\scriptscriptstyle\dagger}_{\mu} b_{\eta+q}
         \rangle_0
         \pm
         \langle  b^{\scriptscriptstyle\dagger}_{\mu} b_{\mu-q}
         \rangle_0
         \langle  
         b^{\scriptscriptstyle\dagger}_{\eta}b_{\eta+q}
         \rangle_0 
         \} 
         \\
         \nonumber
         &=&
        \langle n_\mu  \rangle_0  \pm 
        \langle n_\mu  \rangle_0         \langle n_{\mu -q}  \rangle_0
        +\delta_{q,0}\langle n_\mu  \rangle_0 
         \sum_{\eta} 
         \langle  
        n_{\eta}
         \rangle_0 
         \\
         &=&
        \langle n_\mu  \rangle_0 (1\pm  \langle n_{\mu -q}  \rangle_0)  
        +\delta_{q,0}\langle n_\mu  \rangle_0 
        N
        ,
\end{eqnarray*}
where the $+$ and $-$ signs refer to Bose and Fermi statistics
respectively. Denoting by $ S_{\rm \scriptscriptstyle
  B/F}({{\bf q}},E)$ the dynamic structure factor for a free quantum
gas made up of Bose or Fermi particles one therefore has
\begin{equation}
  \label{202}
  S_{\rm \scriptscriptstyle
  B/F}({{\bf q}},E)=
        {
        1
        \over
         N
        }
         \sum_{\mu}
         \delta
         \left(E-{q^2\over 2m}+ {{{\bf p}}_{\mu}\cdot{{{\bf q}}} \over m}
           \right)
           \langle n_\mu  \rangle_0 (1\pm  \langle n_{\mu -q}  \rangle_0)  
        +\delta_{q,0}\delta (E) N
        .
\end{equation}
The last contribution in (\ref{202}), physically corresponding to
forward scattering (and often neglected in the very definition of
dynamic structure factor) is of no relevance in the present context,
since the contributions corresponding to zero transferred momentum
cancel out in the master-equation, as stressed by the primed sums in
(\ref{pippo}). In the continuum limit (\ref{202}) can be further
simplified evaluating the integral and thus obtaining\cite{art5}
\begin{equation}
\label{arth}
        S_{\rm \scriptscriptstyle B/F}
        ({{\bf q}},E)
        =
        \pm
        {
        1
        \over
         (2\pi\hbar)^3
        }
        {
        2\pi m^2
        \over
        n\beta q
        }
        {
        e^{-\frac{\beta}{2}E}
        \over
        \sinh \left(\frac{\beta}{2}E\right)
        }
        \mathop{\mathrm{arth}}\nolimits
        \left[
        {
        \pm z         
        e^{        -{
        \beta
        \over
             8m
        }
        q^2
        }
         e^{-\frac{\beta}{2}\frac{m}{q^2}E^2}
       \sinh \left( \frac{\beta}{2}E\right)
         \over
         1        \mp z         e^{        -{
        \beta
        \over
             8m
        }
        q^2
        }
        e^{-\frac{\beta}{2}\frac{m}{q^2}E^2}
       \cosh \left( \frac{\beta}{2}E\right)
         }
        \right]  
\end{equation}
or equivalently
\begin{equation}
  \label{abf}
        S_{\rm \scriptscriptstyle B/F}({{\bf q}},E)
        =
        \mp
        {
        1
        \over
         (2\pi\hbar)^3
        }
        {
        2\pi m^2
        \over
        n\beta q
        }
        {
        1
        \over
        1-
        e^{\beta E}
        }
        \log
        \left[
        {
        1\mp z
        \exp{
        \left[
        -{
        \beta
        \over
             8m
        }
        {
        (2mE + q^2)^2
        \over
                  q^2
        }
        \right]
        }
        \over
        1\mp z
        \exp{
        \left[
        -{
        \beta
        \over
             8m
        }
        {
        (2mE - q^2)^2
        \over
                  q^2
        }
        \right]
        }
        }
        \right]
        ,
\end{equation}
where $n$ is the particle density, $\beta$ the inverse temperature and
$z$ the fugacity of the gas, expressed in terms of the chemical
potential $\mu$ by 
$z=e^{\beta\mu}$. For a Bose gas at finite temperature $0\leq z<1$,
while for a Fermi gas $z\geq 0$. If the statistical correction in
(\ref{202}) is left out, so that one is describing a collection of
Maxwell-Boltzmann particles, the expression for the
dynamic structure factor becomes much simpler and is given by
\begin{equation}
  \label{amb}
        S_{\rm \scriptscriptstyle MB}({{\bf q}},E)
        =
        {
        1
        \over
         (2\pi\hbar)^3
        }
        {
        2\pi m^2
        \over
        n\beta q
        }
        z
        \exp\left[
        -{
        \beta m
        \over
             2q^2
        }
        {
        \left(E + \frac{q^2}{2m}\right)^2
        }
        \right]
        .
\end{equation}
Note the presence of the recoil energy in (\ref{amb}), corresponding
to the $\frac{q^2}{2m}$ contribution in the argument of the
exponential, which is essential in order to give the correct dependence on
$E$ in (\ref{sinfinito}), determining the exact factorization property 
(\ref{esatta}) and the operator structure of (\ref{riscoperta}), no
matter how small the recoil energy in each single collision may
actually be.
For more complex systems a direct evaluation of the
dynamic structure factor is of course an extremely complicated task,
nevertheless, even if a direct experimental determination is not
available, approximated or phenomenological expressions can often be
obtained, whose validity can be checked on the basis of general
features such as detailed balance condition and sum
rules.\cite{Lovesey,Griffin}

An important property of the dynamic structure factor, valid in
complete generality, is the fact that it can be expressed as a Fourier
transform, with respect to transferred momentum and energy, of the
time dependent density correlation function\cite{Lovesey} according to
\begin{equation}
  \label{dsf}
  {S} ({\bf q},E)=
        {  
        1  
        \over  
         2\pi\hbar
        }  
        {
        1
        \over
         N
        }
        \int dt 
        {\int d^3 \! {\bf x} \,}        
        e^{
        {
        i
        \over
         \hbar
        }
        (E t -
        {\bf q}\cdot{\bf x})
        }  
        {\int d^3 \! {\bf y} \,}
        \left \langle  
         N({\bf y})  
         N({\bf x}+{\bf y},t)
         \right \rangle
         ,
\end{equation}
a property which as we shall see in Sec.~4.1\ref{structure} will have
an important physical interpretation in the case of the quantum
description of Brownian motion.

\section{Completely Positive Quantum Brownian Motion}
\label{cpqbm}
\noindent
We will now apply (\ref{general}) to the description at quantum level
of Brownian motion,\cite{znf} i.e.\ the dynamics of a massive test particle
interacting through collisions with a fluid of much lighter
particles. This physical model, as a distinguished example of quantum
dissipation, has been the subject of extensive research in the
physical and chemical literature,\cite{Sandulescu,Tannor97,Isar99} and
is still a very debated topic.\cite{art3,reply} On the one hand
Brownian motion is a paradigmatic example of irreversible process and
its correct description at fundamental quantum level should be a
natural gateway to more complex irreversible phenomena; on the other
hand quantum dissipative processes, among which Brownian motion of a
heavy test particle, play a crucial role in many fields of science,
such as nuclear magnetic resonance,\cite{Abragam} quantum
optics\cite{Louisell} and molecular dynamics in condensed
phases,\cite{Weiss,vanKampen} thus leading to a particular interest in
physically and mathematically reliable structures for the description
of quantum dissipation.

We here first consider the case of a fluid made up of
noninteracting Maxwell-Boltzmann particles, where quantum statistics
can be neglected. The relevant dynamic structure factor is given by
(\ref{amb}), where the energy transfer $E$ is actually given by
\begin{equation}
  \label{energia}
  E=E_{p+q} - E_p=        {
        q^2
        \over
             2M
        }
        +
        {{{\bf q}}\cdot{{\bf p}} \over M}
        ,
\end{equation}
with $M$ the mass of the test particle. Writing $S_{\rm
  \scriptscriptstyle MB}({\bf q},E)$ in the form
\begin{displaymath}
        S_{\rm \scriptscriptstyle MB}({\bf q},E)
        =
        {
        1
        \over
         (2\pi\hbar)^3
        }
        {
        2\pi m^2
        \over
        n\beta q
        }
        z
        e^{
        -{
        \beta
        \over
             8m
        }
        q^2
        }
        e^{
        -\frac{\beta}{2}
        E
        }  
        e^{
        -{
        \beta
        \over
             2
        }{
        m
        \over
             q^2
        }
        E^2
        }
\end{displaymath}
one immediately sees that the following identity holds
\begin{equation}
  \label{301}
        S_{\rm \scriptscriptstyle MB}\left({{\bf q}},
        {
        E+E' \over 2
        }\right)
        =
        \sqrt{
        S_{\rm \scriptscriptstyle MB}({{\bf q}},E)
        }
        \sqrt{
        S_{\rm \scriptscriptstyle MB}({{\bf q}},E')
        }
      e^{\frac{\beta}{8}\frac{m}{q^2} (E-E')^2}
      ,
\end{equation}
so that the terms violating the factorization (\ref{x}) are in fact at
least quadratic in the energy difference. Using (\ref{energia})
moreover (\ref{301}) can be written
\begin{equation}
  \label{302}
        S_{\rm \scriptscriptstyle MB}\left({{\bf q}},
        {
        E+E' \over 2
        }\right)
        =
        \sqrt{
        S_{\rm \scriptscriptstyle MB}({{\bf q}},E)
        }
        \sqrt{
        S_{\rm \scriptscriptstyle MB}({{\bf q}},E')
        }
      e^{\frac{\beta}{8m}\frac{\alpha^2}{q^2} [({{\bf p}}-{{\bf
            p}}')\cdot{{\bf q}} ]^2} 
      ,
\end{equation}
where we have denoted by $\alpha=m/M$ the ratio
between the mass $m$ of the particles of the fluid and the mass $M$ of
the test particle. Considering (\ref{pippo}) one now most directly
sees that the term violating the factorization in (\ref{302}) is
indeed negligible provided the statistical operator is
quasi-diagonal. In the Brownian case, when the test particle is much
more massive than the other particles, the factorization is actually
exact, in fact denoting by $ S^{\scriptscriptstyle\infty}_{\rm
  \scriptscriptstyle MB}$ the dynamic structure factor evaluated in
the limit $\alpha\ll 1$ one has
\begin{eqnarray}
  \label{sinfinito}
  \nonumber
        S^{\scriptscriptstyle\infty}_{\rm \scriptscriptstyle MB}
        ({{\bf q}},E)
        &=&
        {
        1
        \over
         (2\pi\hbar)^3
        }
        {
        2\pi m^2
        \over
        n\beta q
        }
        z
        e^{
        -{
        \beta
        \over
             8m
        }
        q^2
        }
        e^{
        -\frac{\beta}{2}
        E
        }
      \\
      &=&
        {
        1
        \over
         (2\pi\hbar)^3
        }
        {
        2\pi m^2
        \over
        n\beta q
        }
        z
        e^{
        -{
        \beta
        \over
             8m
        }
        (1+2\alpha) q^2
        }
        e^{
        -\frac{\beta}{2M} {{\bf q}}\cdot{{\bf p}}
        }
      ,
        \end{eqnarray}
and therefore the dynamic structure factor evaluated in the arithmetic
mean of the energies is exactly equal to the geometric mean of the
dynamic structure factor evaluated in the two different energies
\begin{equation}
  \label{esatta}
        S^{\scriptscriptstyle\infty}_{\rm \scriptscriptstyle
          MB}\left({{\bf q}}, 
        {
        E+E' \over 2
        }\right)
        =
        \sqrt{
        S^{\scriptscriptstyle\infty}_{\rm \scriptscriptstyle MB}({{\bf q}},E)
        }
        \sqrt{
        S^{\scriptscriptstyle\infty}_{\rm \scriptscriptstyle MB}({{\bf q}},E')
        }
      .
\end{equation}
Inserting (\ref{sinfinito}) in (\ref{general}) one has the following
master-equation, first obtained in\cite{art3}
\begin{eqnarray}
        \label{riscoperta}
        \nonumber
        {  
        d {\hat \varrho}  
        \over  
                      dt
        }  
        &=&
        {}-
        {i \over \hbar}
        \left[{
        {\hat {{\sf p}}}^2
        \over
        2M
        }
        ,
        {\hat \varrho}
        \right]
        +
        {2\pi \over\hbar}
        (2\pi\hbar)^3
        n
        \int d^3\!
        {{\bf q}}
        \,  
        {
        | \tilde{t} (q) |^2
        }
        \\
        &&
        \hphantom{cosicosi}
        \times
        \Biggl[
        e^{{i\over\hbar}{{\bf q}}\cdot{\hat {{\sf x}}}}
        \sqrt{
        S^{\scriptscriptstyle\infty}_{\rm \scriptscriptstyle MB}({{\bf
            q}},{\hat {{\sf p}}}) 
        }
        {\hat \varrho}
        \sqrt{
        S^{\scriptscriptstyle\infty}_{\rm \scriptscriptstyle MB}({{\bf
            q}},{\hat {{\sf p}}}) 
        }
        e^{-{i\over\hbar}{{\bf q}}\cdot{\hat {{\sf x}}}}
        -
        \frac 12
        \left \{
        S^{\scriptscriptstyle\infty}_{\rm \scriptscriptstyle MB}({{\bf
            q}},{\hat {{\sf p}}}), 
        {\hat \varrho}
        \right \}
        \Biggr]
        \nonumber
        \\
        &=&
        {}-
        {i \over \hbar}
        \left[{
        {\hat {{\sf p}}}^2
        \over
        2M
        }
        ,
        {\hat \varrho}
        \right]
        +
        z{4\pi^2 m^2 \over\beta\hbar}
        \int d^3\!
        {{\bf q}}
        \,  
        {
        | \tilde{t} (q) |^2
        \over
        q
        }
        e^{-
        {
        \beta
        \over
             8m
        }
        {(1+2\alpha) {{q}}^2}
        }
      \nonumber
        \\
        &&
        \hphantom{cosicosi}
        \times
        \Biggl[
        e^{{i\over\hbar}{{\bf q}}\cdot{\hat {{\sf x}}}}
        e^{-{\beta\over 4M}{{\bf q}}\cdot{\hat {{\sf p}}}}
        {\hat \varrho}
        e^{-{\beta\over 4M}{{\bf q}}\cdot{\hat {{\sf p}}}}
        e^{-{i\over\hbar}{{\bf q}}\cdot{\hat {{\sf x}}}}
        - {1\over 2}
        \left \{
        e^{-{\beta\over 2M}{{\bf q}}\cdot{\hat {{\sf p}}}}
        ,
        {\hat \varrho}
        \right \}
        \Biggr],
\end{eqnarray}
where $z$ for a free gas of Maxwell-Boltzmann particles is given by $n
\lambda_m^3$, with $ \lambda_m=\sqrt{{2\pi \hbar^2\beta}/{m}}$
the thermal de Broglie wavelength of the gas particles. Note that
(\ref{riscoperta}) has in fact the structure of the generator of a
completely positive time evolution.\cite{Lindblad} One can directly
check that an operator of the form
\begin{equation}
  \label{equi}
  \hat{\sf w}_0 (\hat{\sf{p}})=e^{-\beta {
\hat{\sf{p}}^2
\over
     2M
}
}
,
\end{equation}
where $\beta$ is the inverse temperature of the macrosystem and $M$
the mass of the microsystem, is a stationary solution of
(\ref{riscoperta}), in fact
\begin{eqnarray*}
  && \nonumber
        e^{{i\over\hbar}{{\bf q}}\cdot{\hat {{\sf x}}}}
        e^{-{\beta\over 4M}{{\bf q}}\cdot{\hat {{\sf p}}}}
        \hat{\sf w}_0 (\hat{\sf{p}})
        e^{-{\beta\over 4M}{{\bf q}}\cdot{\hat {{\sf p}}}}
        e^{-{i\over\hbar}{{\bf q}}\cdot{\hat {{\sf x}}}}
        - {1\over 2}
        \left \{
        e^{-{\beta\over 2M}{{\bf q}}\cdot{\hat {{\sf p}}}}
        ,
        \hat{\sf w}_0 (\hat{\sf{p}})
        \right \}
        \\
        &&=
         e^{-{\beta\over 2M}{{\bf q}}\cdot ({\hat {{\sf p}}}-{{\bf q}})}
        \hat{\sf w}_0 (\hat{\sf{p}}-{{\bf q}}) - e^{-{\beta\over
            2M}{{\bf q}}\cdot {\hat {{\sf p}}}} 
        \hat{\sf w}_0 (\hat{\sf{p}}) 
        =2\sinh \left({\beta\over 2M}{{\bf q}}\cdot{\hat {{\sf
                p}}}\right) \hat{\sf w}_0 (\hat{\sf{p}})
        ,
\end{eqnarray*}
which is an odd function of ${{\bf q}}$, so that the integral over
the whole space vanishes.

In order to go over from the master-equation
(\ref{riscoperta}) to a Fokker-Planck structure describing quantum
dissipation, corresponding to the quantum description of the classical
Brownian motion,\cite{znf} we consider the limit of small momentum
transfer ${{\bf q}}$, corresponding through the physical interpretation
of the dynamic structure factor to long wavelength fluctuations in the
macrosystem. Expanding the exponentials containing the operators
${\hat {{\sf x}}}$ and ${\hat {{\sf p}}}$ up to second order in
${{\bf q}}$ or equivalently keeping contributions at most bilinear in
${\hat {{\sf x}}}$ and ${\hat {{\sf p}}}$ one has
\begin{eqnarray}
  \label{304}
  && 
  \!\!\!\!\!\!
  \nonumber
          e^{{i\over\hbar}{{\bf q}}\cdot{\hat {{\sf x}}}}
        e^{-{\beta\over 4M}{{\bf q}}\cdot{\hat {{\sf p}}}}
        {\hat \varrho}
        e^{-{\beta\over 4M}{{\bf q}}\cdot{\hat {{\sf p}}}}
        e^{-{i\over\hbar}{{\bf q}}\cdot{\hat {{\sf x}}}}
        - {1\over 2}
        \left \{
        e^{-{\beta\over 2M}{{\bf q}}\cdot{\hat {{\sf p}}}}
        ,
        {\hat \varrho}
        \right \}=
        \\
        && \nonumber
        =
        {}+\frac{i}{\hbar}\sum_{i=1}^3 {{\bf q}}_i[\hat{{\sf x}}_i,
        {\hat \varrho}   ] 
        \\
        && \nonumber
        \hphantom{=}{}
        +\left( \frac{\beta}{4M}\right)^2
        \sum_{i,j=1}^3 {{\bf q}}_i {{\bf q}}_j  \hat{{\sf p}}_i{\hat
          \varrho} \hat{{\sf p}}_j   
        +\frac{1}{2} \left( \frac{\beta}{4M}\right)^2
        \sum_{i,j=1}^3 {{\bf q}}_i {{\bf q}}_j \{ \hat{{\sf p}}_i
        \hat{{\sf p}}_j ,{\hat \varrho}\} 
        \\
        && \nonumber
        \hphantom{=}{}
        -\frac{i}{\hbar} \left( \frac{\beta}{4M}\right) 
         \sum_{i,j=1}^3 {{\bf q}}_i {{\bf q}}_j
         [\hat{{\sf x}}_i,\{\hat{{\sf p}}_j, {\hat \varrho}   \}]
         -\frac{1}{2}\frac{1}{\hbar^2}
         \sum_{i,j=1}^3 {{\bf q}}_i {{\bf q}}_j
         [\hat{{\sf x}}_i,[\hat{{\sf x}}_j, {\hat \varrho}   ]]
        \\
        &&\nonumber 
        \hphantom{=}{}
        -\frac{1}{4} \left( \frac{\beta}{2M}\right)^2
        \sum_{i,j=1}^3 {{\bf q}}_i {{\bf q}}_j \{ \hat{{\sf p}}_i
        \hat{{\sf p}}_j ,{\hat \varrho}\} 
        \\
        && \nonumber
        =
        {}+\frac{i}{\hbar}\sum_{i=1}^3 {{\bf q}}_i[\hat{{\sf x}}_i,
        {\hat \varrho}   ] 
        \\
        &&\nonumber
         \hphantom{=}{}
         -\frac{1}{2}
         \sum_{i,j=1}^3 {{\bf q}}_i {{\bf q}}_j
{         \left\{
         \frac{1}{\hbar^2}
         [\hat{{\sf x}}_i,[\hat{{\sf x}}_j, {\hat \varrho}   ]]
         +
         \left( \frac{\beta}{4M}\right)^2
         [\hat{{\sf p}}_i,[\hat{{\sf p}}_j, {\hat \varrho}   ]]
         +
         \frac{i}{\hbar}  \left(\frac{\beta}{2M} \right)
         [\hat{{\sf x}}_i,\{\hat{{\sf p}}_j, {\hat \varrho}   \}]
         \right\}
}
\\
\end{eqnarray}
and therefore (\ref{riscoperta}) in this limit becomes
\begin{eqnarray}
        \label{305}
        \nonumber
        {  
        d {\hat \varrho}  
        \over  
                      dt
        }  
        =
        &-&
        {i \over \hbar}
        \left[{
        {\hat {{\sf p}}}^2
        \over
        2M
        }
        ,
        {\hat \varrho}
        \right]
        \\
        &-&\nonumber
        z
        {2\pi^2 m^2 \over\beta\hbar}
        \int d^3\!
        {{\bf q}}
        \,  
        {
        | \tilde{t} (q) |^2
        \over
        q
        }
        e^{-
        {
        \beta
        \over
             8m
        }
        {{{q}}^2}
        }
        \sum_{i=1}^3
        {{\bf q}}^2_i
        \times
        \\
        &&
        \biggl\{
        {
        1
        \over
        \hbar^2
        }
        \left[
        {\hat {{\sf x}}}_i ,
        \left[
        {\hat {{\sf x}}}_i , {\hat \varrho}
        \right]
        \right]
        +
        {
        \beta^2
        \over
         16 M^2
        }
        \left[
        {\hat {\sf p}}_i ,
        \left[
        {\hat {\sf p}}_i , {\hat \varrho}
        \right]
        \right]
        +
        {i\over\hbar}
        {
        \beta
        \over
        2M
        }
        \left[
        {\hat {{\sf x}}}_i ,
        \left \{
        {\hat {\sf p}}_i ,{\hat \varrho}
        \right \}
        \right]
        \biggr\}
        ,
\end{eqnarray}
where again because of the integration only terms bilinear in the
momentum transfer and with $i=j$ survive.
Note that the result (\ref{305}) heavily depends on an exact
compensation of the different coefficients in (\ref{304}), leading to
the particularly simple structure of the last line of (\ref{304}),
which is a typical structure of generator of quantum Brownian
motion.\cite{Sandulescu} Supposing without loss of generality the
scattering to be isotropic, we have ${{\bf q}}^2_i = \frac 13 q^2 $,
and the following coefficients related to diffusion and friction can be
introduced:
\begin{eqnarray}
        \label{coeff}  
        D_{pp}
        &=&
        z\frac 23
        {\pi^2 m^2 \over\beta\hbar}
        \int d^3\!
        {{\bf q}}
        \,  
        {
        | \tilde{t} (q) |^2
        }
        q
        e^{-
        {
        \beta
        \over
             8m
        }
        {{{q}}^2}
        }
        \nonumber
        \\
        D_{xx} 
        &=& 
        \left(
        {
        \beta\hbar
        \over
            4M
        }\right)^2 D_{pp}
        \nonumber
        \\
        \gamma 
        &=&
        \left({
        \beta
        \over
             2M
        }\right)
        D_{pp}
        ,
\end{eqnarray}
so that the Fokker-Planck equation (\ref{305}) can be more compactly written
\begin{eqnarray}
        \label{qbm}
        {  
        d {\hat \varrho}  
        \over  
                dt  
        }  
        =
        &-&
        {i\over\hbar}
        [
        {{\hat {{\sf H}}}_0}
        ,{\hat \varrho}
        ]
        -
        {
        D_{pp}  
        \over
         \hbar^2
        }
        \sum_{i=1}^3
        \left[  
        {\hat {{\sf x}}}_i,
        \left[  
        {\hat {{\sf x}}}_i,{\hat \varrho}
        \right]  
        \right]  
        \nonumber \\
        &-&
        {
        D_{xx}
        \over
         \hbar^2
        }
        \sum_{i=1}^3
        \left[  
        {\hat {\sf p}}_i,
        \left[  
        {\hat {\sf p}}_i,{\hat \varrho}
        \right]  
        \right]  
        -
        {i\over\hbar}
        \gamma
        \sum_{i=1}^3
        \left[  
        {\hat {{\sf x}}}_i ,
        \left \{  
        {\hat {\sf p}}_i,{\hat \varrho}
        \right \}  
        \right]      .
\end{eqnarray}
\pagebreak

\subsection{Structural Features}
\label{structure}
\noindent
We now consider some structural features of the mapping giving the
dissipative part of (\ref{qbm}), given by
\begin{equation}
  \label{qdxp}
   {\cal L}^{{\hat {{\sf x}}},
     {\hat {{\sf p}}}}[\hat{\sf w}]\equiv
   {\cal L}[\hat{\sf w}]=
        -
        {
        D_{pp}  
        \over
         \hbar^2
        }
        \sum_{i=1}^3
        \left[  
        {\hat{{\sf x}}}_i ,
        \left[  
        {\hat{{\sf x}}}_i ,\hat{\sf w}
        \right]  
        \right]  
        -
        {
        D_{xx}
        \over
         \hbar^2
        }
        \sum_{i=1}^3
        \left[  
         {\hat{{\sf p}}}_i ,
        \left[  
         {\hat{{\sf p}}}_i ,\hat{\sf w}
        \right]  
        \right]  
        -
        {i\over\hbar}
        \gamma
        \sum_{i=1}^3
        \left[  
        {\hat{{\sf x}}}_i ,
        \left \{  
         {\hat{{\sf p}}}_i ,\hat{\sf w}
        \right \}  
        \right]      
        ,
\end{equation}
where the dependence on the operators ${\hat {{\sf x}}}$ and ${\hat
  {{\sf p}}}$ has been put in major evidence. The mapping ${\cal
  L}^{{\hat {{\sf x}}}, {\hat {{\sf p}}}}$ is said to be covariant
under the action of a unitary representation ${\cal U}_g$ of a
symmetry group $G$ provided the identity
\begin{displaymath}
  {\cal L}^{{\hat {{\sf x}}},{\hat {{\sf p}}}}[{\cal U}_g[\hat{\sf
    w}]]={\cal U}_g 
  [{\cal L}^{{\hat {{\sf x}}},{\hat {{\sf p}}}}[\hat{\sf w}]] 
\end{displaymath}
holds, where
${\cal U}_g [\hat{\sf w}]=\hat{\sf U} (g) \hat{\sf w} \hat{\sf
  U}^{\dagger} (g)$. 
Exploiting the simple relation
\begin{displaymath}
  [\hat{\sf A}, \hat{\sf U}\hat{\sf B}\hat{\sf U}^{\dagger}]_\mp =
 \hat{\sf U}[\hat{\sf U}^{\dagger}\hat{\sf A}\hat{\sf U},\hat{\sf
   B}]_\mp\hat{\sf U}^{\dagger} 
 ,
\end{displaymath}
where 
$\hat{\sf U}$ is
a unitary operator, and considering the explicit structure
(\ref{qdxp}) of ${\cal L}^{{\hat {{\sf x}}},{\hat {{\sf p}}}}$ we have 
\begin{displaymath}
  {\cal L}^{{\hat {{\sf x}}},{\hat {{\sf p}}}}[{\cal U}_g[\hat{\sf
    w}]]={\cal U}_g 
[{\cal L}^{\hat{\sf U}^{\dagger} (g){\hat {{\sf x}}}\hat{\sf U}
  (g),\hat{\sf U}^{\dagger} (g)
{\hat {{\sf p}}}\hat{\sf U} (g)}[\hat{\sf w}]]
,
\end{displaymath}
so that covariance is granted if and only if the condition
\begin{displaymath}
 {\cal L}^{\hat{\sf U}^{\dagger} (g){\hat {{\sf x}}}\hat{\sf U}
  (g),\hat{\sf U}^{\dagger} (g)
{\hat {{\sf p}}}\hat{\sf U} (g)}
= 
 {\cal L}^{{\hat {{\sf x}}},{\hat {{\sf p}}}}
\end{displaymath}
holds.
Let us now consider the symmetry groups of relevance to our physical
context. Under translations the operators position and momentum of the 
particle transform as 
\begin{displaymath}
  \hat{\sf U}^{\dagger} ({{\bf a}}){\hat {{\sf x}}}\hat{\sf U}
  ({{\bf a}})={\hat {{\sf x}}}+{{\bf a}},
  \qquad
  \hat{\sf U}^{\dagger} ({{\bf a}}){\hat {{\sf p}}}\hat{\sf U}
  ({{\bf a}})={\hat {{\sf p}}}
  , 
\end{displaymath}
and one immediately has
\begin{displaymath}
 {\cal L}^{{\hat {{\sf x}}}+{{\bf a}},{\hat {{\sf p}}}}
=
 {\cal L}^{{\hat {{\sf x}}},{\hat {{\sf p}}}}
   .
\end{displaymath}
Considering the group of rotations one has the following
transformation laws
\begin{displaymath}
  \hat{\sf U}^{\dagger} (R){\hat {{\sf x}}}\hat{\sf U}
  (R)=R{\hat {{\sf x}}},
  \qquad
  \hat{\sf U}^{\dagger} (R){\hat {{\sf p}}}\hat{\sf U}
  (R)=R{\hat {{\sf p}}}
  , 
\end{displaymath}
and according to the relation 
\begin{eqnarray*}
&&
\!\!\!\!\!\!\!\!
  \sum_{i=1}^3
        [  
        (R{\hat{\sf u}})_i,
        [  
        (R{\hat{\sf v}})_i,\hat{\sf w}
        ]_\mp  
        ]_\mp
        =
        \sum_{i,j,k=1}^3
        R_{ij}R_{ik}
        [  
        {\hat{\sf u}}_j,
        [  
        {\hat{\sf v}}_k,\hat{\sf w}
        ]_\mp  
        ]_\mp
        \\
        &&
        =
        \sum_{j,k=1}^3
        (R^t R)_{jk}
        [  
        {\hat{\sf u}}_j,
        [  
        {\hat{\sf v}}_k,\hat{\sf w}
        ]_\mp  
        ]_\mp
        =
        \sum_{j=1}^3
        [  
        {\hat{\sf u}}_j,
        [  
        {\hat{\sf v}}_j,\hat{\sf w}
        ]_\mp  
        ]_\mp
\end{eqnarray*}
valid for any couple of vector operators ${\hat {{\sf u}}}$, ${\hat
  {{\sf v}}}$, one still has invariance
\begin{displaymath}
 {\cal L}^{R{\hat {{\sf x}}},R{\hat {{\sf p}}}}
= 
{\cal L}^{{\hat {{\sf x}}},{\hat {{\sf p}}}}
. 
\end{displaymath}
One can also see that an operator with the expected
canonical structure (\ref{equi}) is a stationary solution of (\ref{qbm}) in
that
\begin{displaymath}
{\cal L}[\hat{\sf w}_0 (\hat{\sf{p}})]=0
  ,
\end{displaymath}
due to the relationship
\begin{equation}
  \label{equilibrio}
  \frac{\gamma}{D_{pp}}=\frac{\beta}{2M}
\end{equation}
obeyed by the coefficients defined in (\ref{coeff}). 

The relaxation properties of the dynamics driven by the Fokker-Planck
equation (\ref{qbm}) can be easily obtained considering the adjoint
mapping ${\cal L}'$, which gives the time evolution of the single particle
observables according to:
\begin{eqnarray}
        \label{500}
        {  
        d {\hat {\sf A}}  
        \over  
                dt  
        }  
        =
        &+&
        {i\over\hbar}
        [
        {{\hat {{\sf H}}}_0}
        ,{\hat {\sf A}}
        ]
        + {\cal L}' [{\hat {\sf A}}]
        \nonumber \\
        {}
        =
        &+&
        {i\over\hbar}
        [
        {{\hat {{\sf H}}}_0}
        ,{\hat {\sf A}}
        ]
        -
        {
        D_{pp}  
        \over
         \hbar^2
        }
        \sum_{i=1}^3
        [  
        {\hat {{\sf x}}}_i,
        [  
        {\hat {{\sf x}}}_i,{\hat {\sf A}}
        ]  
        ]  
        \nonumber \\
        &-&
        {
        D_{xx}
        \over
         \hbar^2
        }
        \sum_{i=1}^3
        [  
        {\hat {\sf p}}_i,
        [  
        {\hat {\sf p}}_i,{\hat {\sf A}}
        ]  
        ]  
        +
        {i\over\hbar}
        \gamma
        \sum_{i=1}^3
         \{  
        {\hat {\sf p}}_i,        [  
        {\hat {{\sf x}}}_i ,
        {\hat {\sf A}}        ]
         \}  
      .
\end{eqnarray}
Considering the components of the momentum operator one has, setting
in (\ref{500}) ${\hat {\sf A}}\rightarrow {\hat {\sf p}}_k$
\begin{equation}
  \label{501}
          {  
        d {\hat {\sf p}}_k
        \over  
                dt  
        }  
        = -2\gamma {\hat {\sf p}}_k
        ,
\end{equation}
while for the kinetic energy $\hat{\sf E}={ {\hat {{\sf p}}}^2 \over
  2M }$
\begin{equation}
  \label{502}
          {  
        d {\hat {\sf E}}
        \over  
                dt  
        }  
        = 3 \frac{D_{pp}}{M} -4 \gamma {\hat {\sf E}}
        ,  
\end{equation}
and exploiting (\ref{equilibrio})
\begin{equation}
  \label{503}
          {  
        d {\hat {\sf E}}
        \over  
                dt  
        }  
        = -4\gamma \left({\hat {\sf E}}-\frac{3}{2\beta} \right)
        .
\end{equation}
Due to (\ref{501}), setting $\eta=2\gamma$, the mean value of the
momentum relaxes exponentially to zero, on a typical time
scale $1/\eta$, recovering a result strictly analogous to the
classical one.\cite{Roepke}
Correspondingly (\ref{503}) implies that the mean value of the kinetic
energy of the microsystem reaches for long times the expected
classical value
\begin{displaymath}
  \langle {\hat {\sf E}}\rangle=\frac{3}{2\beta}=\frac{3}{2} {\rm kT},
\end{displaymath}
where T is the temperature of the macrosystem, the relaxation being
also exponential with a rate  $1/2\eta$.

Analogies and differences of (\ref{qbm}) with the classical
Fokker-Planck equations for the description of Brownian motion can be
perhaps most easily seen writing it in terms of the Wigner
function.\cite{Wigner} The Wigner function, related to the statistical
operator by 
the identity
\begin{equation}
  \label{w}
  f_{\rm \scriptscriptstyle W} ({\bf x},{\bf p})=
          \int 
         { d^3\!
        {{\bf k}}
        \over 
        (2\pi\hbar)^3
        }
        \,  
        e^{\frac{i}{\hbar} {\bf x}\cdot{\bf k}} \langle {\bf
          p}+{{\bf k}}/{2}|\hat{\varrho}|{\bf p}-{{\bf
            k}}/{2}\rangle ,
\end{equation}
allows for a phase-space description in quantum
mechanics,\cite{WignerPR} giving a phase-space probability
density.\footnote{The Wigner function defined in (\ref{w}) isn't
  actually a well-defined probability density, since its positivity is
  not granted. Well-defined probability densities can nevertheless be
  introduced in quantum mechanics, exploiting the more modern
  formulation in terms of POV measures.\cite{HolevoOLD,HolevoNEW} Here
  we are only interested in drawing some simple analogies with the
  classical case, so that we will use (\ref{w}) because of its
  simplicity and popularity, together with the appeal of the compact
  expression (\ref{qqdw}).}
Using (\ref{w}) eq.~(\ref{qbm}) becomes
\begin{eqnarray}
  \label{qqdw}
  \frac{\partial}{\partial t}f_{\rm \scriptscriptstyle W} ({\bf
    x},{\bf p})&=& 
  -\frac{{\bf p}}{M}\cdot {\mathbf \nabla}_x 
  f_{\rm \scriptscriptstyle W} ({\bf x},{\bf p})
  + D_{pp} {\bf \Delta}_p
  f_{\rm \scriptscriptstyle W} ({\bf x},{\bf p})
  \nonumber
  \\
  && \hphantom{pippo}
  \nonumber
  \\
  && {}
  + D_{xx} {\bf \Delta}_x
  f_{\rm \scriptscriptstyle W} ({\bf x},{\bf p})
  +2\gamma {\mathbf \nabla}_p \cdot\left({\bf p} f_{\rm \scriptscriptstyle
      W} ({\bf x},{\bf p}) \right)
  ,
\end{eqnarray}
that is to say a Fokker-Planck equation in which, apart from a
dissipative contribution, terms corresponding to diffusion in both
position and momentum appear. The appearance of both these contributions
together is necessary in order that (\ref{qbm}) exhibits a structure
of generator of a completely positive time evolution,\cite{art3} and
therefore appears as a peculiar quantum feature. 

Many other similar examples of Fokker-Planck equations describing
dissipation can be found in the literature (for a review
see\cite{Sandulescu,Tannor97,Isar99}), mainly relying on
phenomenological approaches or on a characterization of the formal
structures compatible with complete positivity. They differ in the
definition of the coefficients, in their relative weight, and in the
presence or absence of various dissipative contributions or of
potential terms which determine the underlying free dynamics. All
these factors determine the general properties such as existence of a
stationary solution, invariance under symmetry transformations,
complete positivity.\cite{reply} This properties for (\ref{qbm}) and
more generally for (\ref{general}) are considered in some detail
in.\cite{art5} In particular a distinguishing feature of
(\ref{general}) is the existence of a canonical stationary solution
provided the state of the macrosystem with which the microsystem is
interacting is a $\beta$-KMS state.\cite{KMS}

The unique feature of the Fokker-Planck equation (\ref{qbm}) from a
physical point of view is its derivation as a long wavelength limit of
the master-equation (\ref{general}), which gives the subdynamics of
the microsystem in terms of the dynamic structure factor of the
medium: this corresponds to a Kramers-Moyal expansion in the small
parameter ${\bf q}$ characterizing the size of the
fluctuations.\cite{vanKampen,Risken} This connection between
(\ref{general}) and (\ref{qbm}) gives a profound justification for the
selection of possible dissipative contributions appearing in
(\ref{qbm}) and for the exact expressions of the coefficients in
(\ref{coeff}). More than this, observing that the dynamic structure
factor can also be expressed as a Fourier transform, with respect to
transferred momentum and energy, of the time dependent density
correlation function according to (\ref{dsf}), one has a direct
physical connection between Brownian motion of the test particle and
density fluctuations in the medium. This physical intuition was in
fact the starting point of Einstein's approach to the classical
description of Brownian motion, his key idea being that the random
motion was due to the discrete nature of matter.\cite{Einstein}
Eq.~(\ref{qbm}) also has a mathematically distinguishing feature,
since it can be written in explicit Lindblad form in terms of a single
generator for each Cartesian direction, a feature also indicated by
the general structure of translation covariant quantum dynamical
semigroups obtained by Holevo,\cite{HolevoTI} one of the few
characterizations of quantum dynamical semigroups with unbounded
generators.\cite{HolevoLNPH} In fact introducing the operators
\begin{displaymath}
  {\hat {{\sf a}}}_i=
{
\sqrt{2}
\over
 \lambda_M
}
\left(
{\hat {{\sf x}}}_i
+{i\over\hbar}
{
\lambda_M^2
\over
                    4
}
{\hat {{\sf p}}}_i
\right),
\end{displaymath}
where $ \lambda_{M}=\sqrt{\hbar^2\beta / M} $ and $[ {\hat
  {{\sf a}}}_i , {\hat {{\sf a}}}_j^{\scriptscriptstyle\dagger} ]
=\delta_{ij}$, (\ref{qbm}) can be also written:
\begin{eqnarray}
        \label{generator}
        {  
        d {\hat \varrho}  
        \over  
        dt
        }  
        =  
        &-&
        {i\over\hbar}
        [
        {{\hat {{\sf H}}}_0}
        ,{\hat \varrho}
        ]
        -
        {
        D_{pp}  
        \over
         \hbar^2
        }
        {
        \lambda_M^2
        \over
                 4
        }
        \sum_{i=1}^3
        \left[  
          {\hat   {{\sf a}}}_i{}^2 - {\hat
            {{\sf a}}}_i^{\scriptscriptstyle\dagger}{^2}       ,{\hat
            \varrho} 
        \right]
        \nonumber
        \\
        &+&
        {
        D_{pp}
        \over
         \hbar^2
        }
        \lambda_M^2
        \sum_{i=1}^3
        \left[  
        {
        {\hat {{\sf a}}}_i
        {\hat \varrho}
        {\hat {{\sf a}}}_i^{\scriptscriptstyle\dagger}
        - {\scriptstyle {1\over 2}}
        \{
        {\hat {{\sf a}}}_i^{\scriptscriptstyle\dagger}
        {\hat {{\sf a}}}_i
        ,{\hat \varrho}
        \}
        }
        \right]      .
\end{eqnarray} 


\subsection{Quantum Statistics}
\label{qs}
\noindent
As a last application of (\ref{general}) we briefly sketch the
completely new results one obtains when considering the Brownian limit
$\alpha\ll 1$ in the case of a free Bose or Fermi gas,\cite{art4,art5}
starting from expression (\ref{abf}) for the dynamic structure factor
which takes compactly both statistics into account. A suitable
expansion of (\ref{abf}), despite being much more complicated than in
the case of Maxwell-Boltzmann particles leads to a structurally
similar result. Apart from the linear dependence on the fugacity $z$
in the coefficients defined in (\ref{coeff}) the dissipative part of
(\ref{qbm}) is now multiplied by an overall factor
\begin{displaymath}
  \frac{1}{1-z}
\end{displaymath}
for Bose particles and
\begin{displaymath}
  \frac{1}{1+z}
\end{displaymath}
for Fermi particles, so that one has
\begin{eqnarray}
        \label{qbmb}
        {  
        d {\hat \varrho}  
        \over  
                dt  
        }  
        =
        &-&
        {i\over\hbar}
        [
        {{\hat {{\sf H}}}_0}
        ,{\hat \varrho}
        ]
        -
        \frac{1}{1-z}
        \left\{
        {
        D_{pp}  
        \over
         \hbar^2
        }
        \sum_{i=1}^3
        \left[  
        {\hat {{\sf x}}}_i,
        \left[  
        {\hat {{\sf x}}}_i,{\hat \varrho}
        \right]  
        \right]  
        \right.
        \nonumber \\
        &+&
        \left.
        {
        D_{xx}
        \over
         \hbar^2
        }
        \sum_{i=1}^3
        \left[  
        {\hat {\sf p}}_i,
        \left[  
        {\hat {\sf p}}_i,{\hat \varrho}
        \right]  
        \right]  
        +
        {i\over\hbar}
        \gamma
        \sum_{i=1}^3
        \left[  
        {\hat {{\sf x}}}_i ,
        \left \{  
        {\hat {\sf p}}_i,{\hat \varrho}
        \right \}  
        \right]  
        \right \}    
\end{eqnarray}
for the Fokker-Planck equation describing Brownian motion at quantum
level in a free gas of Bose particles and
\begin{eqnarray}
        \label{qbmf}
        {  
        d {\hat \varrho}  
        \over  
                dt  
        }  
        =
        &-&
        {i\over\hbar}
        [
        {{\hat {{\sf H}}}_0}
        ,{\hat \varrho}
        ]
        -
        \frac{1}{1+z}
        \left\{
        {
        D_{pp}  
        \over
         \hbar^2
        }
        \sum_{i=1}^3
        \left[  
        {\hat {{\sf x}}}_i,
        \left[  
        {\hat {{\sf x}}}_i,{\hat \varrho}
        \right]  
        \right]  
        \right.
        \nonumber \\
        &+&
        \left.
        {
        D_{xx}
        \over
         \hbar^2
        }
        \sum_{i=1}^3
        \left[  
        {\hat {\sf p}}_i,
        \left[  
        {\hat {\sf p}}_i,{\hat \varrho}
        \right]  
        \right]  
        +
        {i\over\hbar}
        \gamma
        \sum_{i=1}^3
        \left[  
        {\hat {{\sf x}}}_i ,
        \left \{  
        {\hat {\sf p}}_i,{\hat \varrho}
        \right \}  
        \right]  
        \right \}    
\end{eqnarray}
for the corresponding dynamics in a free gas of Fermi
particles. Comparing (\ref{qbm}) and (\ref{qbmb}) or (\ref{qbmf})
one immediately sees (recalling that the fugacity is a positive
number, restricted to be less than one for Bose particles) that the
friction coefficient $\gamma$ for Maxwell-Boltzmann particles given by
the last contribution in (\ref{coeff}) is now enhanced to
\begin{displaymath}
  \frac{\gamma}{1-z}
\end{displaymath}
in the case of Bose particles and suppressed to
\begin{displaymath}
  \frac{\gamma}{1+z}
\end{displaymath}
in the case of Fermi particles.

\section{Conclusions and Outlook}
\label{end}
\noindent
In this paper we have presented some new results in the description
of the subdynamics of a microsystem interacting through collisions
with a macrosystem that have recently been obtained, especially in
connection with quantum dissipation and the quantum description of
Brownian motion. These results rely on a recent general approach to
the formulation of subdynamics in a non-relativistic field theoretical
framework, where a major emphasis has been put on the choice of a
suitable subset of relevant observables, whose subdynamics is to be
studied on a time scale over which they are slowly varying. Particular
attention is given to structural properties of the mapping giving the
reduced irreversible evolution, which should satisfy
complete positivity or some less stringent version of this property.

The main result presented is the master-equation (\ref{general})
describing the interaction of a probe particle with some macroscopic
system in which the subdynamics is driven by the dynamic structure
factor of the system, a two-point correlation function whose physical
features are considered in Sec.~3.1\ref{structure-factor}, embodying its
statistical mechanics properties.  Considering the long wavelength
limit of this master-equation the Fokker-Planck equation (\ref{qbm})
for the description of Brownian motion is obtained, in which the
quantum statistics of the fluid can be taken into account, leading to
(\ref{qbmb}) and (\ref{qbmf}). These corrections due to quantum
statistics to the description of quantum dissipation have been
introduced for the first time and deserve further investigation, since
they lead in principle to experimentally observable effects. 
Apart form applications in the study of quantum dissipation and
decoherence of a particle interacting with the surrounding medium
(with regard to the problem of decoherence considered in this
framework see also\cite{gargnano99-torun01}),
these results might be used for the study of motion of test particles
in degenerate quantum gases. Degenerate samples of dilute, weakly
interacting Bose and Fermi atoms have in fact recently been
experimentally realized,\cite{bec,fermi} showing in particular the
phenomenon of Bose-Einstein condensation, which is now the object of
intense study (see\cite{rmp,nc} for a review). Also expressions for
the dynamic structure factor, which take interactions into account and
improve the result for the free case, have been recently introduced
and studied.\cite{stringari}

A natural extension of the formalism within this approach is of course
the application to many-body macroscopic systems, obtaining quantum
kinetic equations for the subdynamics of suitable subsets of slowly
varying observables, and will be the object of future research work.

\nonumsection{Acknowledgements}
\noindent
This work was financially supported by MURST under
Cofinanziamento and Progetto Giovani.

\nonumsection{References}

\end{document}